\documentclass{aa}

\usepackage{graphicx,epsfig,fancyhdr,psfig,rotating,amsmath,natbib}

\def\kms{\rm km\;s$^{-1}$}
\def\etal{{et al. }}

\begin{document}

\title{Dynamics and plasma properties of an X-ray jet  from SUMER, EIS, XRT and EUVI A \& B simultaneous observations }

\author{M.S. Madjarska}

\offprints{madj@arm.ac.uk}
\institute{Armagh Observatory, College Hill, Armagh BT61 9DG, N. Ireland}

 \date{Received date, accepted date}

\abstract
{Small-scale transient phenomena in the quiet Sun are believed to play an important role in coronal heating and solar wind generation. One of them named as ``X-ray jet'' is the subject of our study.}
{We indent to investigate the dynamics, evolution and physical properties of this phenomenon.}
{ We combine spatially and temporally multi-instrument  observations obtained  simultaneously with the SUMER spectrometer onboard SoHO, EIS and XRT onboard  Hinode,  and  EUVI/SECCHI  onboard the Ahead and Behind  STEREO  spacecrafts. We derive plasma parameters such as   temperatures and densities as well as  dynamics by using spectral lines formed in  the temperature range from 10\,000~K to 12~MK. We also use image difference technique to investigate the evolution of the complex  structure of the studied phenomenon.}
{With the available unique combination of data we were able to establish that the formation of a jet-like event is triggered by not one but several energy depositions which  are most probably originating from magnetic reconnection. Each energy deposition is followed by the expulsion of pre-existing or new reconnected loops and/or collimated flow along open magnetic field lines. We derived in great detail the dynamic process of X-ray jet formation and evolution. We also found for the first time  spectroscopically in the quiet Sun a temperature of 12~MK and density of 4~$\times$~10$^{10}$~cm$^{-3}$ in a reconnection site.  We raise an issue  concerning an uncertainty in using the SUMER Mg~{\sc x} 624.9~\AA\ line for coronal diagnostics.  We clearly identified two types of up-flow: one collimated up-flow along open magnetic field lines and a plasma cloud formed from the expelled  BP loops. We also report a cooler down-flow along closed magnetic field lines. A comparison is made with a model developed by Moreno-Insertis \etal\ (2008). }
{}

\authorrunning{Madjarska, M. S.}
\titlerunning{Properties of X-ray jets}

\maketitle

\section{Introduction}

X-ray jets in the solar atmosphere were first  seen in images from the Soft X-ray Telescope 
\citep[SXT]{1991SoPh..136...37T} onboard the Yohkoh satellite. They  appear as collimated flows 
originating from coronal bright points (BPs) or active regions and are always associated with
transient brightenings named as micro-flares. We will review  here  only  the very recent studies on these
phenomena. \citet[][and the references therein]{2007PASJ...59S.745S} resolved the fine structure of a X-ray jet and reported  that a
loop appeared near the footpoint of the jet when a footpoint brightening was observed. They also
found that the X-ray jet appears after the loop breaks. They observed thread like structures
along the jet.  \citet{2007PASJ...59S.751C} found from  Extreme-ultraviolet Imaging Spectrometer (EIS) 
40\arcsec\ slot observations of a polar coronal hole that  jet temperatures range from 0.4~{\rm MK} to 5.0~{\rm MK}.
 The  jet velocities had values which are less than the escape solar velocity ($\approx$ 618 \kms). 
They interpreted the increase of the radiance in the cooler spectral lines 
as emitted from the falling back plasma of the cooling down jet.  \citet{nishi08} reported on simultaneous 
cool emission  ($\sim$10$^4$~K) in  Ca~{\sc ii}~H  images from  the Solar Optical Telescope onboard Hinode 
and hot emission (1 $\times$ 10$^6$~K)  in Fe~{\sc xii}~195~\AA\ images from the Extreme-ultraviolet Imaging Telescope (EIT) onboard the 
Solar and Heliospheric Observatory (SoHO) as well as in X-ray Telescope (XRT)  images (Al\_poly filter with temperature
response with a maximum at 5 $\times$ 10$^6$~K)  from a  `giant' jet. From their observations of a current-sheet-like structure seen in
all three instruments, they concluded that magnetic reconnection is occurring in the transition region or upper chromosphere.   \citet{2007Sci...318.1580C} found that X-ray jets 
 have two distinct velocities: one near the Alfv\'en speed ($\sim$800~\kms) and another near the sound speed (200 ~\kms).  \cite{2008ApJ...680L..73P} discussed the first stereoscopic observations of 
polar coronal jets made by the Extreme-ultraviolet imagers  (EUVI) of the SECCHI instruments onboard the twin 
STEREO spacecrafts. The authors depicted  a helical structure of a jet which showed signs  of untwisting
with the jet initially ascending slowly with $\approx$ 10--20~\kms\ and then suddenly accelerating to velocities larger than 300~\kms. Helical structures in jets have  also been reported by \citet{1996PASJ...48..123S, 2007A&A...469..331J,2009SoPh..259...87N}.  \citet{2010A&A...510L...1K} studied the relation of a so-called macrospicules in He~{\sc ii}~304~\AA\ and an X-ray jet. The Doppler shifts of the jet and the inclination angle of the jet were ``attributed to a rotating motion of the macrospicule rather than a radial flow or an expansion''. The authors concluded that the macrospicule is driven by the unfolding motion of a twisted magnetic flux rope, while the associated X-ray jet is a
radial outflow.

It is believed that the majority of the jet-like phenomena in the solar atmosphere are produced by magnetic reconnection.  Two-dimensional  \citep{1977ApJ...216..123H,  1995Natur.375...42Y, 1996PASJ...48..353Y,2007ApJ...657L..53I}, 2.5 dimensional \citep{1998ApJ...495..491K, 2007A&A...466..367A} as well as  three-dimensional  (3D) \citep{2008ApJ...673L.211M, 2009ApJ...691...61P}   magneto-hydodynamic (MHD) models  have been developed. 

The feature analysed here was selected from a large number of jet-like events registered during a multi-instrument observing campaign. We exploit here the very rare circumstance that such a jet-like event has simultaneously been registered by 
two spectrometers and several imagers. Due to the complexity of both the analysis and the amount of information contained in the data, we decided to present this study in an individual article, while a statistical study on the dynamics and physical parameters of jet-like events will follow in a separate publication. Sect. 2 contains detailed information on all analysed observations including the alignment, both spatially and temporally, of the different type of observations.  Sect.~3 presents  the analysis and gives the results on the dynamics, temperature and density of the event. The obtained physical parameters and dynamics compared to already published similar information, and a  discussion  on the observable characteristics of jet-like phenomenon  in the light of a theoretical 3D MHD model by \citet{2008ApJ...673L.211M}  are given in Sect.~4.

\section{Observations}
\label{sect2}
\begin{figure}[!h]
\begin{center}
\includegraphics [scale=1.]{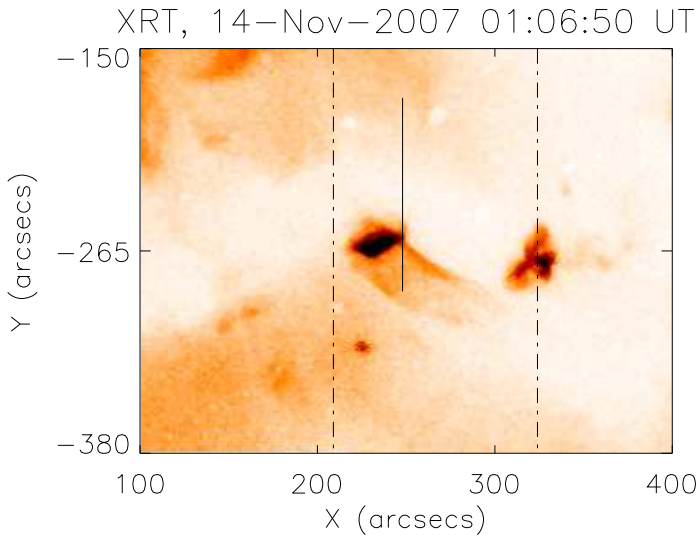}
\caption{Al\_poly XRT image (color table reversed) of the jet-like feature studied here. The vertical
continuous line corresponds to the position of the SUMER slit. The dashed-dotted line denotes the
horizontal and part of the vertical field-of-view of the EIS raster.}
\label{fig1}
\end{center}
\end{figure}

The event discussed here occurred at the boundaries of an equatorial hole on 2007 November 14 
(Fig.~\ref{fig1}).  It was registered with the  Solar Ultraviolet  Measurements of Emitted Radiation  (SUMER) 
spectrometer onboard SoHO,  EIS and XRT on board Hinode, and the EUVI in the Fe~{\sc ix/x} 171~\AA\ channel  on board the A and B STEREO  spacecrafts during a specially planned
 multi-instrument observing campaign.    The  field-of-views (FOVs) of XRT, EIS and SUMER are shown 
 in Fig.~\ref{fig1}.    Details on  the registered SUMER spectral lines can be found in Table~\ref{T1}. 
 The analysed EIS spectral  lines are listed in Table~\ref{T2}.  No magnetic field data exist for the observed event.

\begin{table}
\centering
\caption{SUMER spectral lines, The expression `/2' means that the spectral
 line was observed in second order. The comment `b' denotes that the spectral line is 
 blending a  close-by line.}
\label{T1}
\begin{tabular}{c c c c c}
\hline\hline
Ion & Wavelength(\AA) & logT$_{max}$/K\\
\hline
N~{\sc v} & 1238.82 & 5.3 \\
C~{\sc i} & 1248.00 & 4.0 \\
& 1248.88 & &\\
C~{\sc i} & 1249.00 & 4.0 \\
O~{\sc iv}/2 & 1249.24b & 5.2 \\ 
Si~{\sc x}/2 & 1249.40b & 6.1  \\
C~{\sc i} & 1249.41 & 4.0 \\     
Mg~{\sc x}/2 & 1249.90 & 6.1  \\
Si~{\sc ii} & 1250.09&4.1   \\
O~{\sc iv}/2 &1250.25b&5.2\\
Si~{\sc ii} & 1250.41 &4.1  \\
C~{\sc i} & 1250.42b & 4.0 \\
S~{\sc ii} & 1250.58 &4.2 \\
Si~{\sc ii} & 1251.16 &4.1  \\
C~{\sc i} & 1251.17 & 4.0 \\
Si~{\sc i} & 1256.49&4.0\\
Si~{\sc i} & 1258.78 &4.0 \\
S~{\sc ii} & 1259.53b &4.2 \\
O~{\sc v}/2&1259.54&5.4  \\
Si~{\sc ii}&1260.44&4.1\\
\hline
\end{tabular}
\end{table}

\begin{table}
\centering
\caption{EIS observed spectral lines. The comment `b' indicates  that the spectral line is 
blended \citep[see][]{2007PASJ...59S.857Y, 2008ApJS..176..511B, 2009A&A...495..587Y}.}
\begin{tabular}{c c c c c}
\hline\hline
Ion & Wavelength(\AA) & logT$_{max}$/K \\
\hline
He~{\sc i} & 256.32b & 4.7 \\
O~{\sc v} & 192.90 & 5.4\\
O~{\sc vi} & 184.12 & 5.5\\
Mg~{\sc vi}& 270.39b & 5.7\\ 
Mg~{\sc vii} & 278.39 & 5.8\\
& 280.75 & 5.8 & \\     
Si~{\sc vii} & 275.35 & 5.8 \\
Fe~{\sc viii} &185.21&5.8\\
Fe~{\sc x} & 184.54&6.0  \\
Fe~{\sc xi} & 188.23 &6.1 \\
Si~{\sc x} & 258.37 &6.1 \\
 & 261.04 &6.1  \\
Fe ~{\sc xii} & 195.12 &6.1\\
Fe~{\sc xiii}&202.04&6.2\\
&203.82&6.2\\
Fe~{\sc xiv} &274.20b &6.3 \\
Fe~{\sc xv} & 284.16 &6.3 \\
Fe~{\sc xvi}&262.98&6.4&\\
Ar~{\sc xv}&293.69&6.6\\
Fe~{\sc xxiii}&263.76&7.1\\
\hline
\end{tabular}
\label{T2}
\end{table}

\subsection{SUMER}
The SUMER \citep{1995SoPh..162..189W,
1997SoPh..170..105L} observations were made  from 01:01~UT until 02:56~UT during a dedicated multi-instrument 
observing  campaign targeting coronal hole boundaries. A slit with a size of 
1\arcsec$\times$300\arcsec\ was used. The detector B was exposed for 60~{\rm s} while the spectrometer was 
observing in a sit-and-stare mode. The time sequence was followed by a raster with a
size of 60\arcsec $\times$ 300\arcsec\  made only in the O~{\sc v}~629.73 \AA\ and  
Si~{\sc i}~1256.49~\AA\ lines. The observations were compensated for the solar rotation. Five 
spectral  windows were transferred to the ground  each with a size of 50 spectral pixels $\times$ 300 spatial pixels. 
The spectral lines read-outs are shown in Table~\ref{T1}. From all lines only  O~{\sc v}~629.73~\AA\ 
was taken on the bare part of the detector. At the start of the observations (at 01:01~UT) the 
spectrometer was pointed at solar disk coordinates  xcen~=~-217~\arcsec\  and 
ycen~=~257\arcsec.

\subsection{TRACE}
During the  TRACE  \citep{1999SoPh..187..229H} observations, the imager was already in
an eclipse period.  That is why we selected the three ultraviolet  filters  at 1550~\AA,
1600~\AA, and 1700~\AA\  which are not  influenced by the absorption of the Earth
atmosphere. Images in the Fe~{\sc ix/x}~171~\AA\ filter were taken with a cadence of 30~{\rm min} and were used 
 for co-alignment purposes. The observations started at 03:27,  i.e.
 $\approx$3 hrs later than SUMER and Hinode. The three UV channel observations were obtained with a 
 cadence of 10~{\rm s}. The FOV of the images is 256\arcsec~$\times$~256\arcsec\ with a 
spatial resolution of 1\arcsec.

\subsection{EIS}

The EIS \citep{2007SoPh..243...19C} study was specially designed to provide the best possible 
coverage of spectral lines with formation temperatures  (logT$_{max}$) from 4.7 to 7.1~K
(Table~\ref{T2}). First, a raster with a size of 120\arcsec $\times$ 512\arcsec\ was obtained 
followed by 50 rasters with a size of 24\arcsec  $\times$ 512\arcsec starting at 00:20~UT. 
The feature analysed in this work was registered only in the big raster covering the entire area of the jet from the start to 
the end of the eruption (Fig.~\ref{fig1}). An exposure time of  60~{\rm s} and 2\arcsec\
slit were used during these observations.


\begin{figure*}
\includegraphics{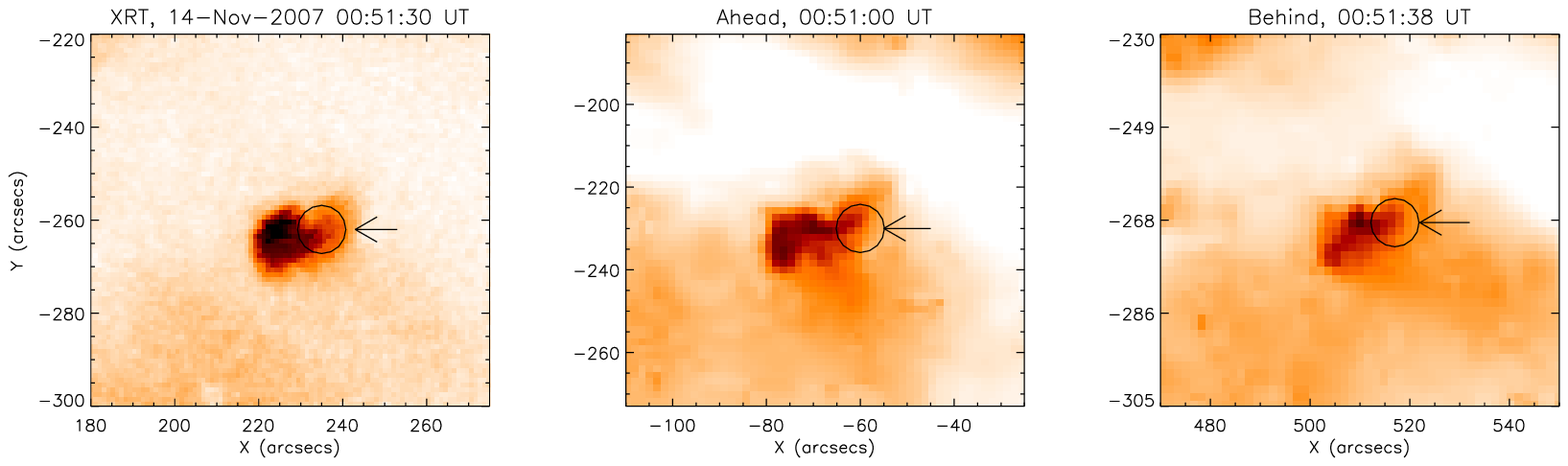}
\includegraphics{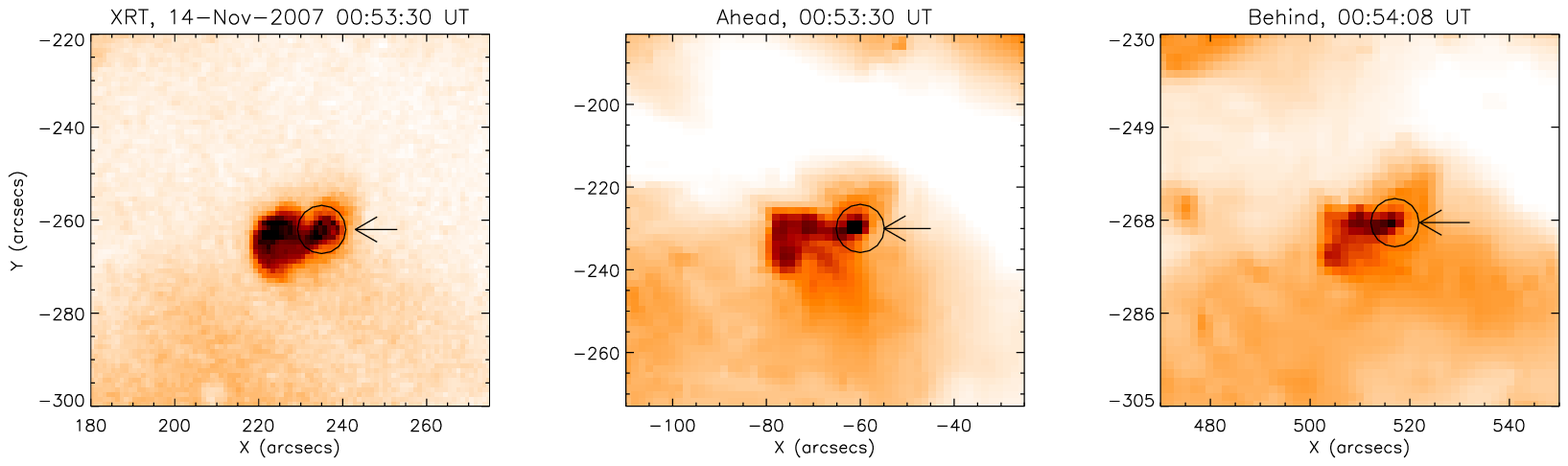}
\caption{{\bf From left to right:} XRT, EUVI A and B images. The top row shows the images before the first micro-flare took place while the bottom row present the images taken seconds after it happened.  The arrows point at the region where the micro-flare (surrounded by a circle) occurred.}
\label{fig2}
\end{figure*}

\subsection{XRT}
 
 The XRT~\citep{2007SoPh..243...63G} was observing with the Al\_poly filter in a FOV of 
 384\arcsec~$\times$~384\arcsec. The exposure time was 16.38~{\rm s} and the cadence 40~{\rm s}.

\subsection{EUVI/SECCHI}
EUVI/SECHHI data from both the Behind (B) and Ahead (A) spacecrafts of the STEREO mission were used 
in the present study taken between 00:20~UT and 2:00~UT . The cadence of EUVI A was variable but the exposure time 
was constantly 4~{\rm s} with the exception of 8 
images (out of 82) where the exposure time was 16~{\rm s}. EUVI B recorded only 47 images in the above mentioned 
period with variable cadence and constant exposure time of approximately  4~{\rm s} with one exception of 16~{\rm s}.

\subsection{Data reduction and co-alignment of the different instruments}

The data from all instruments were reduced using standard software. Only for SUMER the 
standard data reduction  software could not  provide  a satisfactory level of data corrections for 
various instrumental effects. We applied additional corrections for the flatfield and geometrical distortion.

All images were de-rotated to a reference time of 02:27~UT which was a good compromise between the
data taken with different instruments. The co-alignment of 
the  instruments was based on  aligning structures seen at similar temperatures. We used a 
TRACE~171~\AA\ image  taken at 03:27~UT and co-aligned it  with an XRT image taken at 
02:30~UT. The Al\_poly filter has a peak response  at logT= 6.9~K but it can also register emission at temperatures as low as  
logT= 5.5~K. A few bright coronal structures which showed little changes during 2h30m of
observations  were co-aligned. Next an XRT image of the peak of the jet evolution was aligned with 
the  XRT image from 02:30~UT. Knowing the internal offsets of the TRACE channels we 
selected a TRACE 1550~\AA\ image taken at the same time of the event as the TRACE~171~\AA\ image already 
aligned with  XRT. Then the TRACE~1500~\AA\ was co-aligned with the SUMER O~{\sc v}~629.73~\AA\ and
Si~{\sc i}~1256.49~\AA\ raster images. Finally, the EIS raster was co-aligned with all above mentioned instruments by using
spectral lines with the best suitable formation temperatures. The  precision of the alignment is 1\arcsec--2\arcsec.  In Fig.~{\ref{fig0} (exist only as online material) the images 
used for alignment are shown. Note that the different appearance of some structures is due to the time difference of the obtained images. Nevertheless, there exist long living structures which remain almost unchanged during the observations. The final adjustment was made from the comparison of the blue-shifted emission of EIS He~{\sc ii} 256.32~\AA\ with the SUMER O~{\sc v}~629.73~\AA\ and N~{\sc v}~1238.82~\AA. 
Due to the dynamic nature of the observed phenomenon,  the combination of spectroscopic and imager data
provided a straightforward alignment of the STEREO data with the rest of the observations. }


\begin{figure*}
\hspace{-5cm}
\includegraphics[scale=1.4]{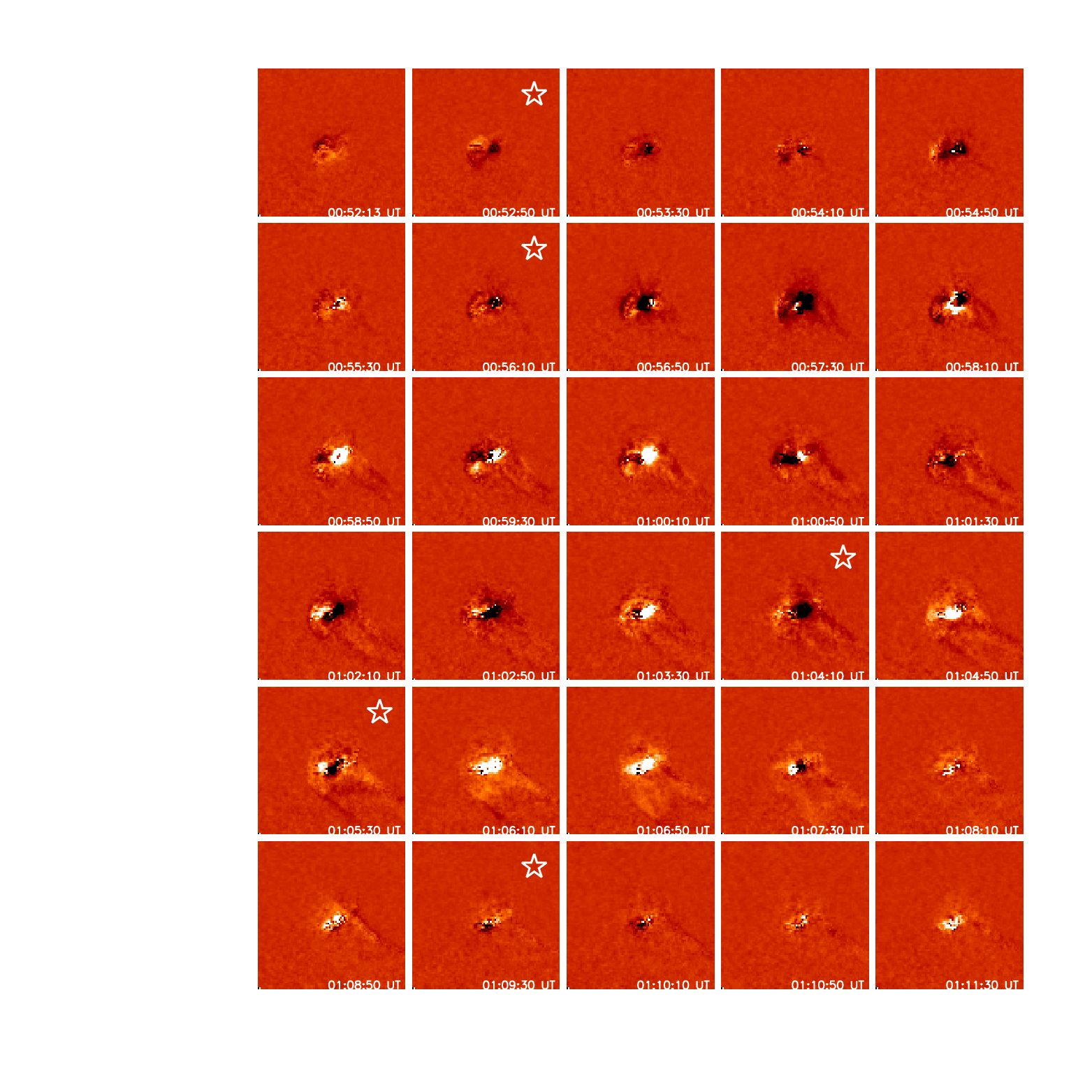}\\

\caption{Sequence of XRT difference images with over-plotted starts which indicate the moments of energy release.  }
\label{fig3}
\end{figure*}


\begin{figure*}
\center
\includegraphics[scale=1.]{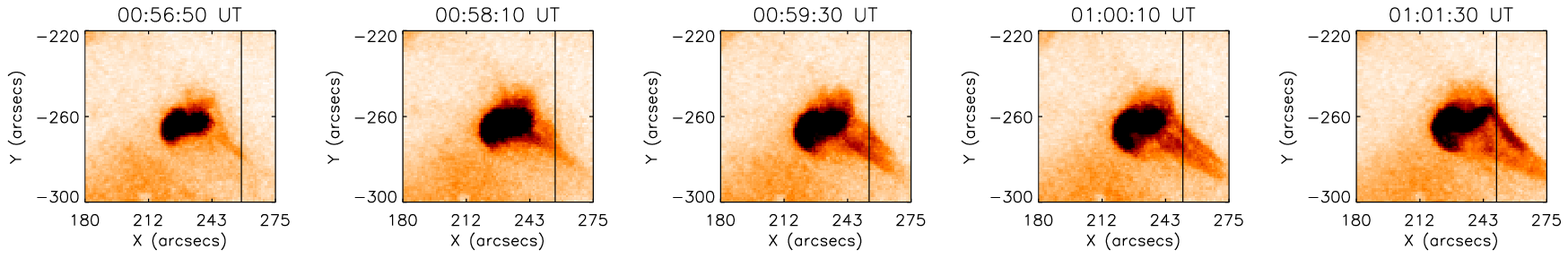}\\
\includegraphics[scale=1.]{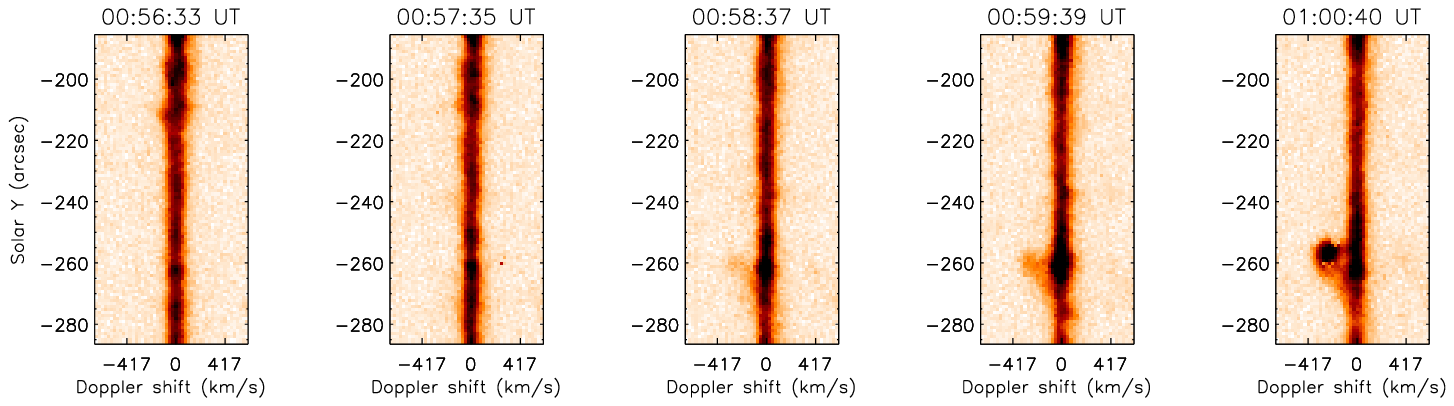}\\
\includegraphics[scale=1.]{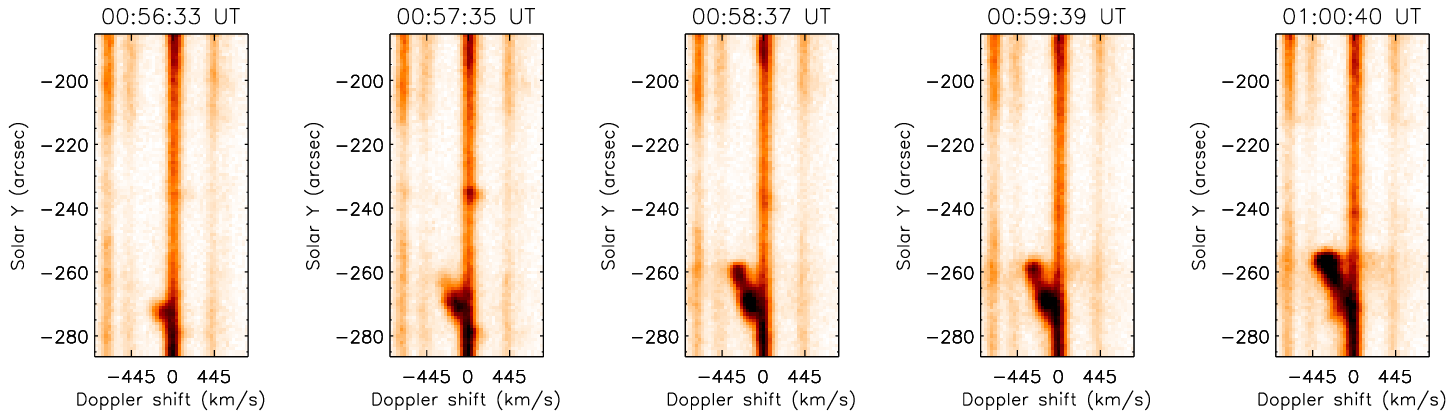}
\caption{XRT images  (top) of the jet-like event shortly after it started. EIS He~{\sc ii} 256.32~\AA\ (middle) and Fe~{\sc xii} 195.12~\AA\  (bottom) spectra along the EIS slit registered around the time the XRT images were taken. Note that the timing of these images is before EIS reaches the SUMER slit position. The slit moves in West--East with a step of 2\arcsec\  at each exposure.}
\label{fig4}
\end{figure*}


\begin{figure*}[!ht]
\center
\includegraphics{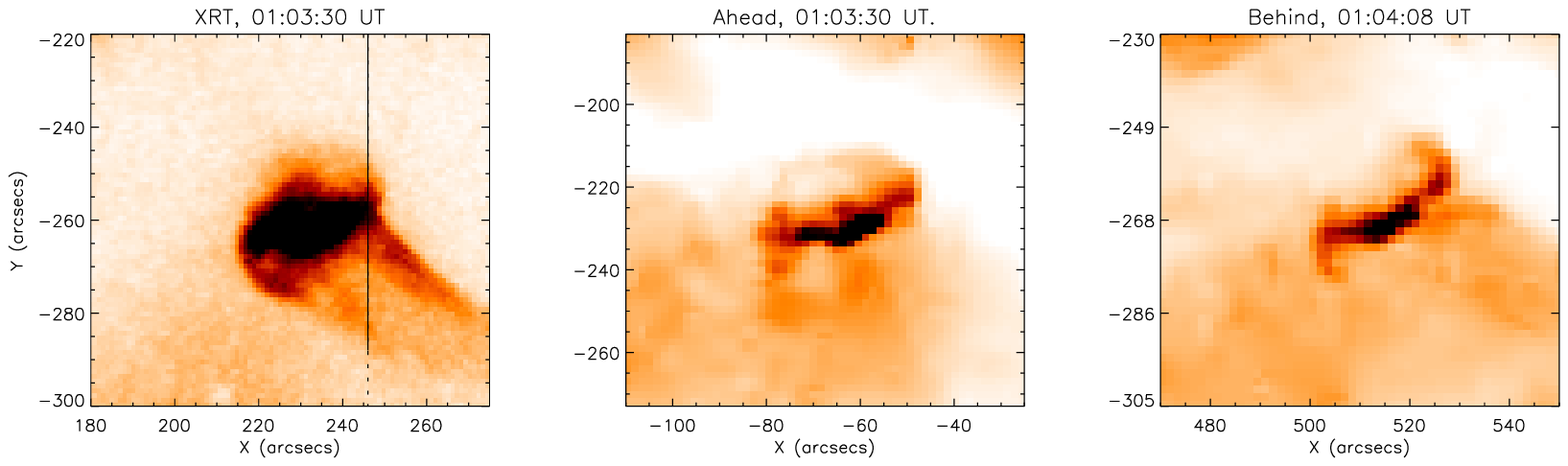}
\includegraphics[scale=1.]{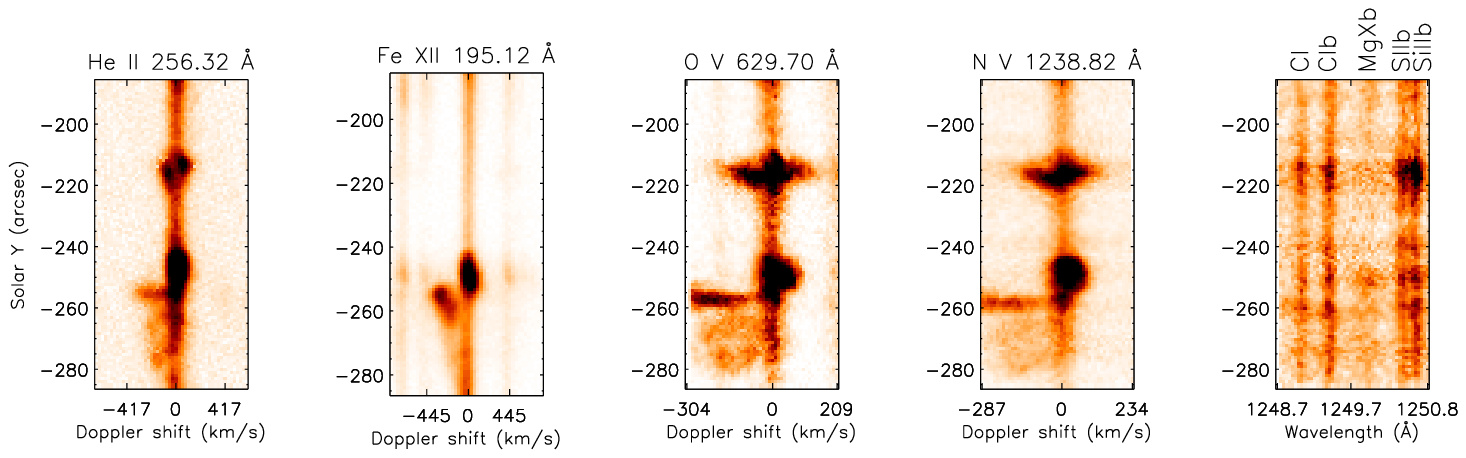}
\caption{{\bf Top:} XRT image with over-plotted   SUMER slit position and EUVI A and  B images during the jet-like phenomenon. 
 {\bf Bottom (from left to right):} EIS He~{\sc ii} 256.32~\AA\ and Fe~{\sc xii} 195.12~\AA\   spectra along the EIS slit at 01:02:44~UT; SUMER O~{\sc v} 
and N~{\sc v} spectra along the SUMER slit; and a spectral window comprising C~{\sc i} 1248.0~\AA, C~{\sc i} 1249.0~\AA, 
Mg~{\sc x} 624.9~\AA\ taken in second order, Si~{\sc ii} 1250.41~\AA\ and S~{\sc ii} 1250.58~\AA\ (for details on blends and 
formation temperatures consult Table~\ref{T1}) at 01:02:019~UT.}
\label{fig5}
\end{figure*}

\section{Data analysis and results}

The phenomenon presented  here was detected in the course of a  statistical study of brightening events in the quiet Sun 
and coronal holes \citep{2010arXiv1002.1675S}.  \citet{2010arXiv1002.1675S} showed that numerous brightenings occur along coronal hole boundaries and  inside coronal holes as  most of them were visually identified as  jet-like features from pre-existing coronal bright points. Their spatial appearance suggested that  they are a product of magnetic reconnection between  the open   magnetic field lines of coronal holes  and the closed ones of pre-existing  magnetic loops named coronal bright points. 

The distinctiveness of our analysis lies in the use of  the so far best combination of imager  and spectroscopic data which were taken simultaneously and cover the entire lifetime of a jet-like  feature.  The X-ray telescope  was monitoring an equatorial
 coronal hole region together with the surrounding quiet Sun during several hours on 2007 November 14
 (for more information on the data see Sect.~\ref{sect2}). The EIS slit was scanning the area in the XRT field-of-view in
  west--east direction crossing the boundary of an equatorial coronal hole while the SUMER slit was in a sit-and-stare mode and positioned in respect to the EIS field-of-view as  shown in Fig.~\ref{fig1}.  The SUMER and EIS slits were pointing at roughly the same position at around 01:02~UT  (no more than 1\arcsec\--2\arcsec\ offset or none).

\subsection{Dynamics analysis and results}

We produced an animated sequence of XRT images (see the online movie) which undoubtedly revealed that the 
studied  feature  (Fig.~\ref{fig1})  is a typical example  of a so-called X-ray jet from a coronal BP  \citep{1992PASJ...44L.173S}.   The event  started with a sudden brightening in an area of  3-4 arcsec$^2$ size in a pre-existing coronal bright point  at 00:52:50~UT and  was registered by all three imagers -- XRT, EUVI A and B  (Fig.~\ref{fig2}). We identified this region  as the so-called micro-flare region where the energy deposition took place.  After producing  a sequence of difference images (Fig.~\ref{fig3}) by subtracting from each image the precedent one, we were surprised to discover that an energy deposition, most probably from magnetic reconnection, happened several times during the event.  We were able to identify 5 bursts during around 16~{\rm min}. This can be seen in Fig.~\ref{fig3} on the images marked with stars.  We should note that although more darkenings (a sudden brightening corresponds to a darkening in a  difference image) are seen, many of them come from  plasma flows as determined by the spectroscopic observations (read later in the text) which  create  in the difference images bright and dark patches.  The  energy deposition seems to occur in different places. 

The BP appeared enlarged (in 3 dimensions corresponding to an expansion) on the XRT images, 40~{\rm s}  after the first energy deposition took place, while loops are clearly seen to expand  4~{\rm min} later, i.e. around 00:56:50~UT, after the second and, as it seems  larger energy deposition (Fig.~\ref{fig3}). The first clearly distinguishable expanding loops can be seen in the XRT difference image around 00:57~UT in Fig.~\ref{fig3} and  the online movie. Following the event in EUVI A (B as well but at lower cadence) images we can clearly see that a large number of loops took off from the BP forming a plasma cloud which moved away from the source region (see the EUVI A image sequence in the online movie, after 01:03~UT).

   In Fig.~\ref{fig4} we show the spectra of the He~{\sc ii}~256.34~\AA\ and Fe~{\sc xii}~195.12~\AA\ lines obtained from 00:56:33~UT, i.e. {4 min after the energy deposition took place. } The top row in the figure displays the XRT images taken approximately in the middle of the EIS exposure duration. The corresponding EIS slit positions are over-plotted.   At  00:56:33~UT is  the first registration of the  jet-like event in EIS as blue-shifted emission. Note that SUMER was not observing at this time yet.  The expanding plasma is only seen in the high temperature lines as, for instance, Fe~{\sc  xii}  (up to Fe~{\sc xv}).  No indication of cooler plasma is seen  until 00:57:35~UT in Si~{\sc vii} (logT$_{max}$= 5.8~K) and until 00:58:37~UT in He~{\sc ii} (logT$_{max}$= 4.7~K, note that the blue wing of He~{\sc ii} is free from blends).   After comparing this observation with the animation sequence,  we can speculate on two possible scenarios: one is that  the hotter and taller loops of the BP started first to expand and  later were followed by the lower lying cooler loops 2 min later. The other scenario is   that the plasma from the energy deposition site is first propelled along the  newly reconnected open magnetic field lines. The latter is consistent with the observation that a strong plasma outflow was registered along the straight magnetic field lines situated in the north-west side of the jet-like event as seen in projection on the disk. There is also the possibility  that these elongated  structures represent  loop legs which became quasi-open, i.e they  are not open into the interplanetary space.   As can be seen in Fig.~\ref{fig4}, in 4 mins the plasma was accelerated from 130~\kms\  to 310~\kms\ derived from a double Gauss-fit of the Fe~{\sc xii} line.  The Doppler shifts of the escaping loops are reaching up to 400--450 \kms (blue-shifted emission). No falling-back plasma  (red-shifted emission)  was observed in the location of the jet.

 We present the observations of XRT, EUVI/SECCHI, SUMER and EIS obtained  as close as possible in time in a composite figure (Fig.~\ref{fig5}). The timing  of the data  corresponds to the moment when both spectrometers were staring at the same position on the solar disk at around 01:02~UT.  The same combination of data for another six pointing times are available in the electronic edition of the journal (Figs. 8 to 13). Note that in spectrometers the time of a spectrum is defined as the time of the beginning of the exposure,  while the time of an image is defined with the end of the exposure.
The appearance of the event  in all three imagers is quite different due to their different view angles. This multi-dimensional view of the studied phenomenon is of great importance for the understanding of the  spectroscopic imprints of the studied event.  

From the spectroscopic observations we established two main flows:  a very fast outflow which regenerates after each energy deposition and a down-flow guided along curved (most probably closed)  magnetic field lines (best seen in EUVI B). Since SUMER started observing only at 01:01:19~UT, the dynamics  in the transition region O~{\sc v} 629~\AA\ and N~{\sc v} 1238~\AA\ lines was caught after this time with strong  blue-shifted emission which expands beyond the spectral windows.  Two sources seem to produce these Doppler shifts as judged from their appearance (Fig.~\ref{fig5}):  one confined in a small area of 2\arcsec--3\arcsec\  producing a strong emission which we associated with the north-west ray-like feature, and the other, more diffuse, cloud like emission, which could have been produced by the surrounding plasma pushed away by the  shock generated from the energy deposition site or/and the expanding loops. 

 The energy deposition 01:04~UT produced  fast rising loops  (Fig.~\ref{fig2a}) which generate Doppler shifts of up to 250~{\kms}. It is clearly noticeable that the SUMER slit cuts through the legs of these rising loops. The emission from the coronal line yet again displays an additional line shift which reflects the plasma stream  along  open field lines seen  in the XRT image. The picture of the outflowing loops persist in the next two exposures at 01:04~UT and 01:05~UT, shown in Figs.~\ref{fig3a} and \ref{fig4a}. The hot outflow is not anymore seen in EIS as the EIS slit moves away from it (it has to be kept in mind that the EIS slit was scanning the area, while the SUMER slit was in  a sit-and-stare position). After the last energy deposition at 01:05:30~UT, an outflow (the strongest in3 terms of emissivity) is observed only along the north-west part of the event.

 The continuous down-flow shows Doppler shifts of up to 25--50~\kms\ in the SUMER transition region lines and up to 150~\kms\ in the EIS coronal lines (Fig.~\ref{fig5a}).  As the EIS slit is moving across   the energy deposition  site,  the red-shift persists, although at 01:06~UT a blue-shifted component in the He~{\sc ii} line is well distinguishable. From the available data we cannot speculate where this blue-shifted emission  comes from.

 The lifetime of the entire process of jet formation and evolution was approximately 27 min from the moment the first reconnection was registered until a jet-like structure fully disappears in the imagers. The dynamic phase of the jet from the moment it was first registered as an outflow until no signature of outflow is seen in the SUMER spectral lines is 18 min. The down-flow  persisted  2--3 min longer.
 
  In Fig. ~\ref{fig8} we present the line profiles of several spectral lines. The profiles were produced from the jet region showing very high velocity.  We selected only spectral lines which are free from closeby spectral lines on the blue side of the spectrum.   The He~{\sc ii} line is free of blends but it is very possible that the blending hot lines have some contribution to the emission of the He~{\sc ii} and its wing. The O~{\sc vi} line at 184.12~\AA\ is also producing  strong blue-shifted emission but this emission is blended by  another O~{\sc vi} line  at 183.94~\AA. The later line is usually up to two times weaker than O~{\sc vi} 184.12~\AA\ and therefore the emission in excess would come from the blue wing of  O~{\sc vi} 184.12~\AA. The comparison of the spectral lines Fe~{\sc viii},~{\sc x},~{\sc xii} and~{\sc xii} shows  an  increase of the emission with the increase of the formation temperature of the spectral line suggesting that the amount of hotter plasma dominates the outflow.  The Doppler shift of the emitted plasma estimated from a double Gauss fit is similar, e.g.  246~\kms, 259~\kms, 279~\kms and 265~\kms for Fe~{\sc viii},~{\sc x},~{\sc xii} and~{\sc xii}, respectively.  Unfortunately, the low signal-to-noise ratio of the higher ionisation lines does not permit to investigate their line profiles in this region.

\subsection{Temperature analysis and results}

\begin{figure*}[!ht]
\center

\vspace{10cm}
\includegraphics{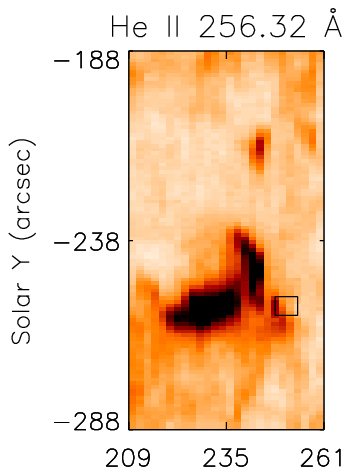}
\includegraphics{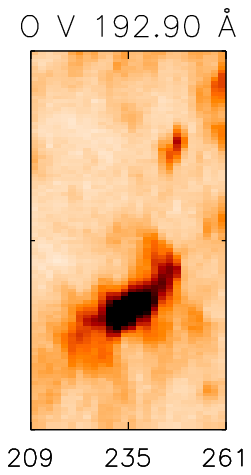}
\includegraphics{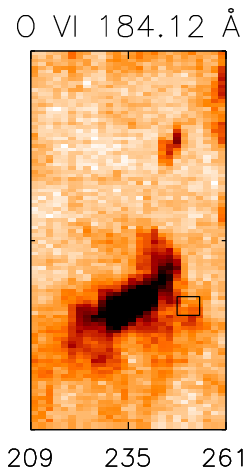}
\includegraphics{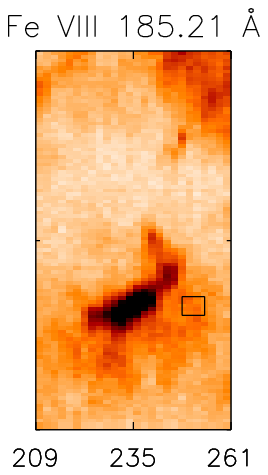}
\includegraphics{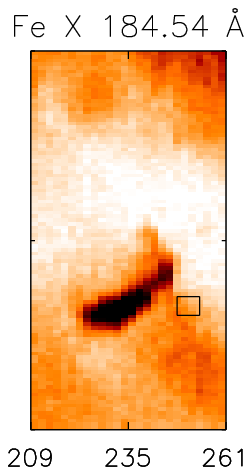}
\includegraphics{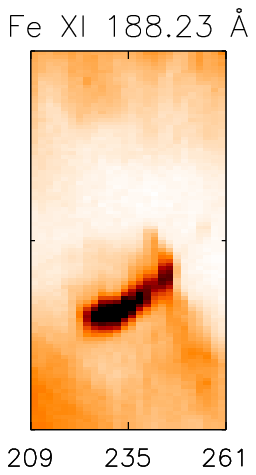}
\includegraphics{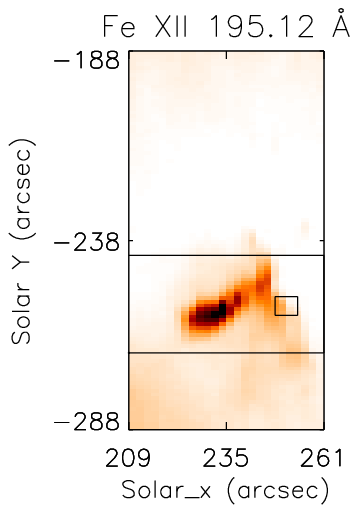}
\includegraphics{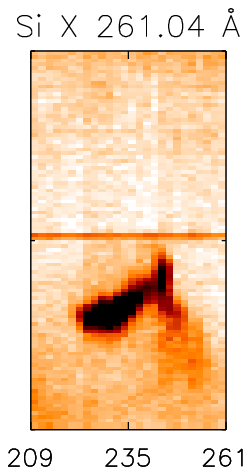}
\includegraphics{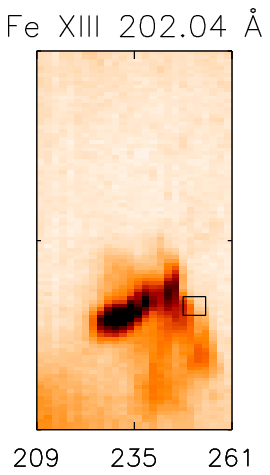}
\includegraphics{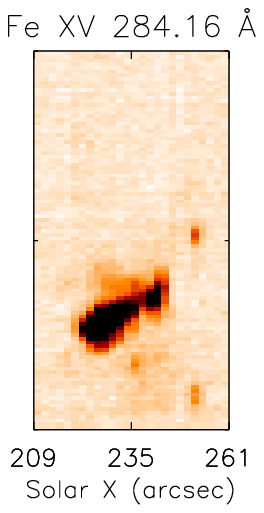}
\includegraphics{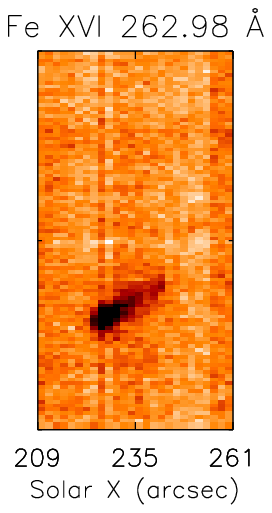}
\includegraphics{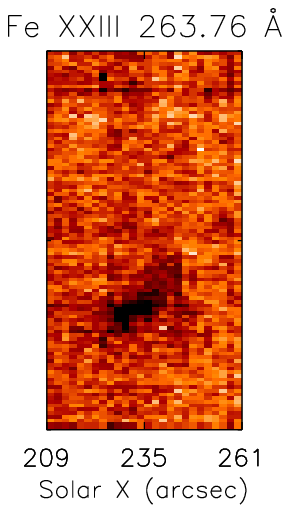}

\caption{The EIS rasters taken in 12 spectral lines with formation temperatures ranging from  50\,000~K to 12~MK  assuming ionisation equilibrium. The over-plotted box of the Fe~{\sc xii} raster denotes the field-of-view shown in Figure~\ref{fig7}. Details on the formation temperature of each line can be found in Table~\ref{T2}.}
\label{fig6}
\end{figure*}

\begin{figure*}[!t]
\center

\vspace{10cm}
\includegraphics{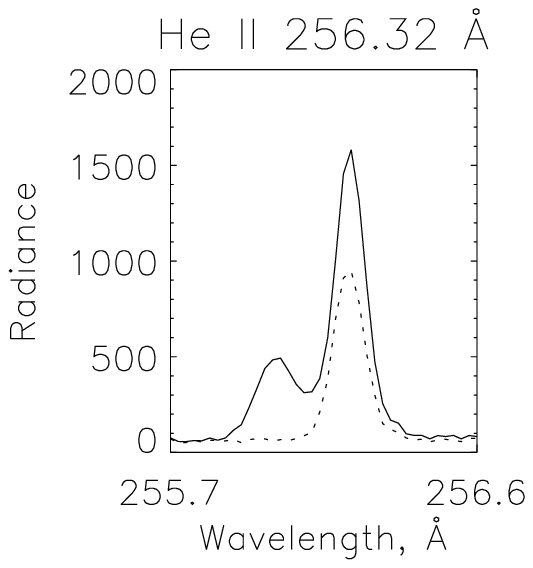}
\includegraphics{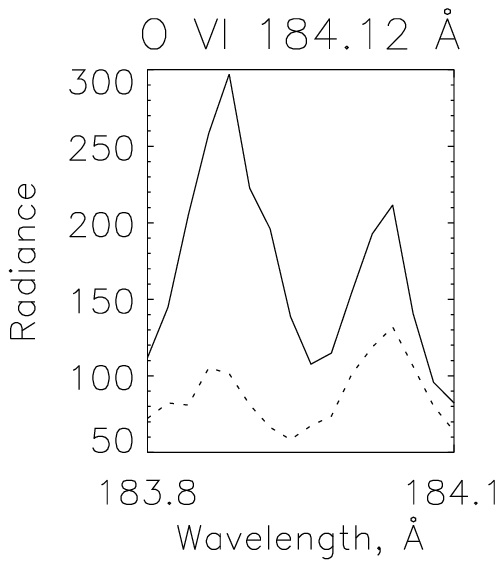}
\includegraphics{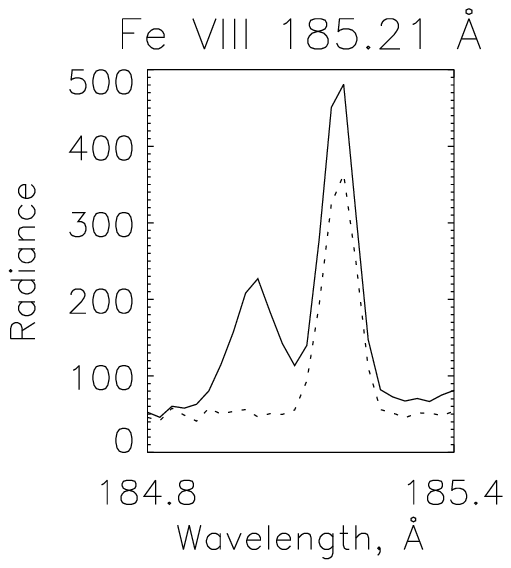}

\includegraphics{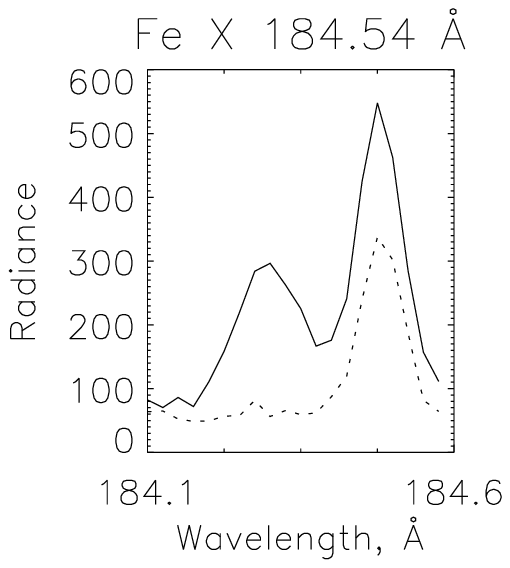}
\includegraphics{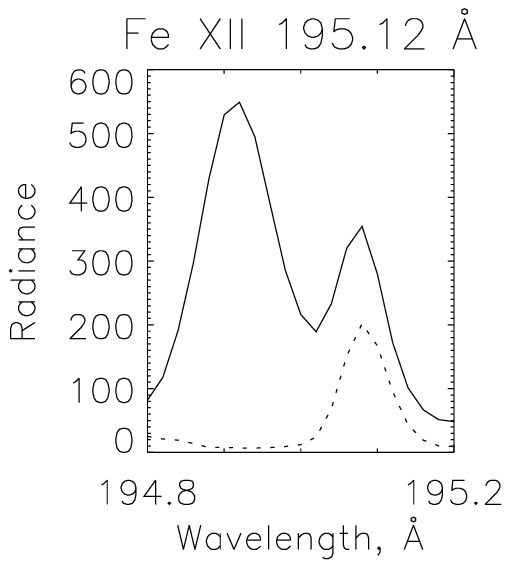}
\includegraphics{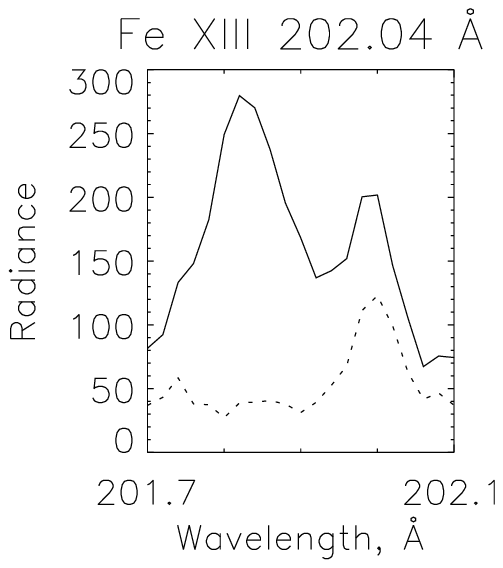}

\caption{ EIS line profiles taken from the region shown with a small box in Fig.~\ref{fig6}. The radiance is in units erg cm$^{-1}$ s sr$^{-1}$ \AA. The dotted line corresponds to a reference profile.}
\label{fig8}
\end{figure*}

 That brings us to the temperature analysis of the emitting plasmas. We can separate the jet into three main temperature regions: the energy deposition site, the up-flowing (blue-shifted emission) plasma of the jet-like event and the down-flow (red-shifted emission). The micro-flaring site emits in a large temperature  range from 0.5~MK (He~{\sc ii}) to up to 12~MK (Fe~{\sc xxiii}  263.76~\AA) (Fig.~\ref{fig6}).  Actually, for the first time  a plasma temperature of 12~MK was detected in a quiet Sun region.  The Fe~{\sc xxiii}  263.76~\AA\ line was  first identified  in observations with the Naval Research Laboratory spectrograph  during a two-ribbon flare by \citet{1975ApJ...197L..33W}. It was also reported by \citet{2008A&A...481L..69D} in a flare observed with EIS.  On the blue side of the line is Ar~{\sc xv}~263.69~\AA\ which can easily be  separated from Fe~{\sc xxiii}   with a double Gauss-fit.

The outflowing plasma has a temperature from 0.5~MK (He~{\sc ii}) to  2~MK (Fe~{\sc xv}) as already reported by \citet{2007PASJ...59S.751C}. The authors used slot observations which carry some uncertainties because of line blends in the observed spectral windows. However, there exist a possibility of even higher temperature plasma which may not be detectable in EIS observations due to the weakness of the higher ionisation lines like for instance Fe~{\sc xvi} and Fe~{\sc xvii}.  No signature was detected in the SUMER chromospheric lines which are formed at temperatures as low as 10\,000~K (C~{\sc i}; see Table~\ref{T1} for more details).

The third region which corresponds to the down-flowing plasma seen as red-shifted emission has temperatures ranging from 15\,000~K (S~{\sc ii}) to 1.3~MK (Fe~{\sc xii} and Si~{\sc x}).  That clearly suggests a cooler plasma downstream.

We need to raise an issue concerning an uncertainty in using the SUMER Mg~{\sc x} 624.9~\AA\ line for coronal diagnostics. If we compare the simultaneous EIS and SUMER observations shown in Fig.~\ref{fig5}, it is clearly noticeable the very weak response of Mg~{\sc x} while Fe~{\sc xii}  which is  formed at a similar temperature has a very strong increase. The Mg~{\sc x} response in the outflow is fully absent  and shows only a very modest increase in the energy deposition region. In the past this spectral  line was used to determine coronal response of transient features in the quiet Sun with a main conclusion that most of these phenomena do not reach coronal temperatures due to the lack of emission in the Mg~{\sc x} line. This, consequently, had put under  question their direct contribution to coronal heating. The  count rate  in the Mg~{\sc x} 624.9 ~\AA\ line as observed in second order of diffraction  is 0.17 counts/s and if we assume a minimum 10\% contribution from the blending Si~{\sc ii} 1250.09~\AA\ and another few percent from other lines (see for more details \citealt{2002A&A...392..309T}) we arrive at  0.15 counts/s. In comparison, the count rate for the Fe~{\sc xii} line is 6.17 counts/s which gives a ratio of the two lines (Fe~{\sc xii}/ Mg~{\sc x}) of around 40.   How can this discrepancy 
be explained? \citet{1989SoPh..122..245G} found that a possible reason for the lower variability of coronal emission in  Mg~{\sc x}  can be due to very slow ionisation and recombination time scales. Although, it may take Mg~{\sc x} several hundred seconds to reach ionisation equilibrium (assuming N$_e$ =10$^9$~cm$^{-3}$),
Mg~{\sc x}  should still be very strong at the 1.5--2~MK even in transient ionisation
conditions (J.G. Doyle, private  communication). We investigated the Mg~{\sc x}  emissivity using the atomic database CHIANTI v6.0 with ionisation fractions from \citet{1998A&AS..133..403M}, we found  that for a  quiet Sun DEM (differential emission emasure) and active region DEM (if we assume a bright point to be a mini active region) the Mg~{\sc x} line is approximately six times weaker than Fe~{\sc xii}. That still does not  explain the lack of counts in the  Mg~{\sc x} line. The other possible reason  is a low SUMER sensitivity  for observations in second order of diffraction. 

\subsection{Density analysis and results}

\begin{figure}[!h]
\hspace{-2cm}
\includegraphics[scale=0.65]{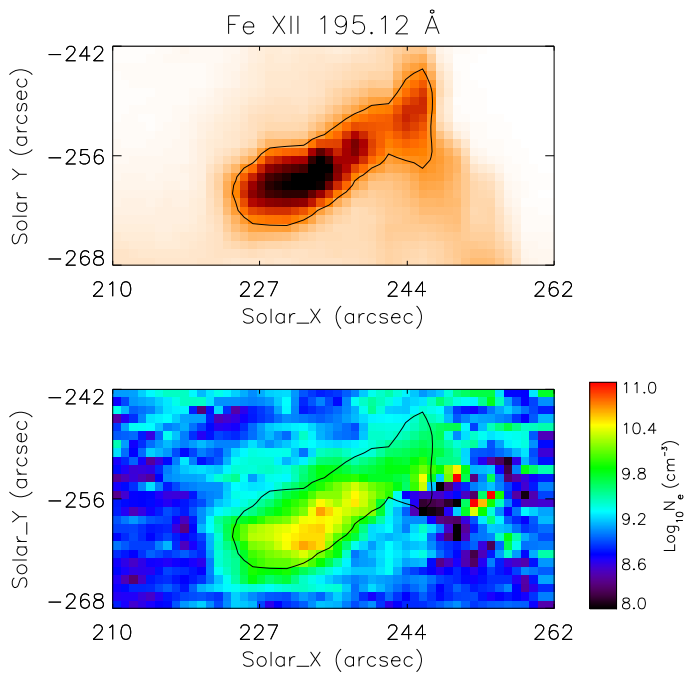}
\caption{{\bf Top:} A partial view of the radiance image in the Fe~{\sc xii} line shown in Figure~\ref{fig6}. {\bf Bottom:} An electron density image derived from the Fe~{\sc xii} line ratio (see Sect.~3.3 for more details).}
\label{fig7}
\end{figure}

\begin{figure*}[!h]
\center
\includegraphics[scale=0.7]{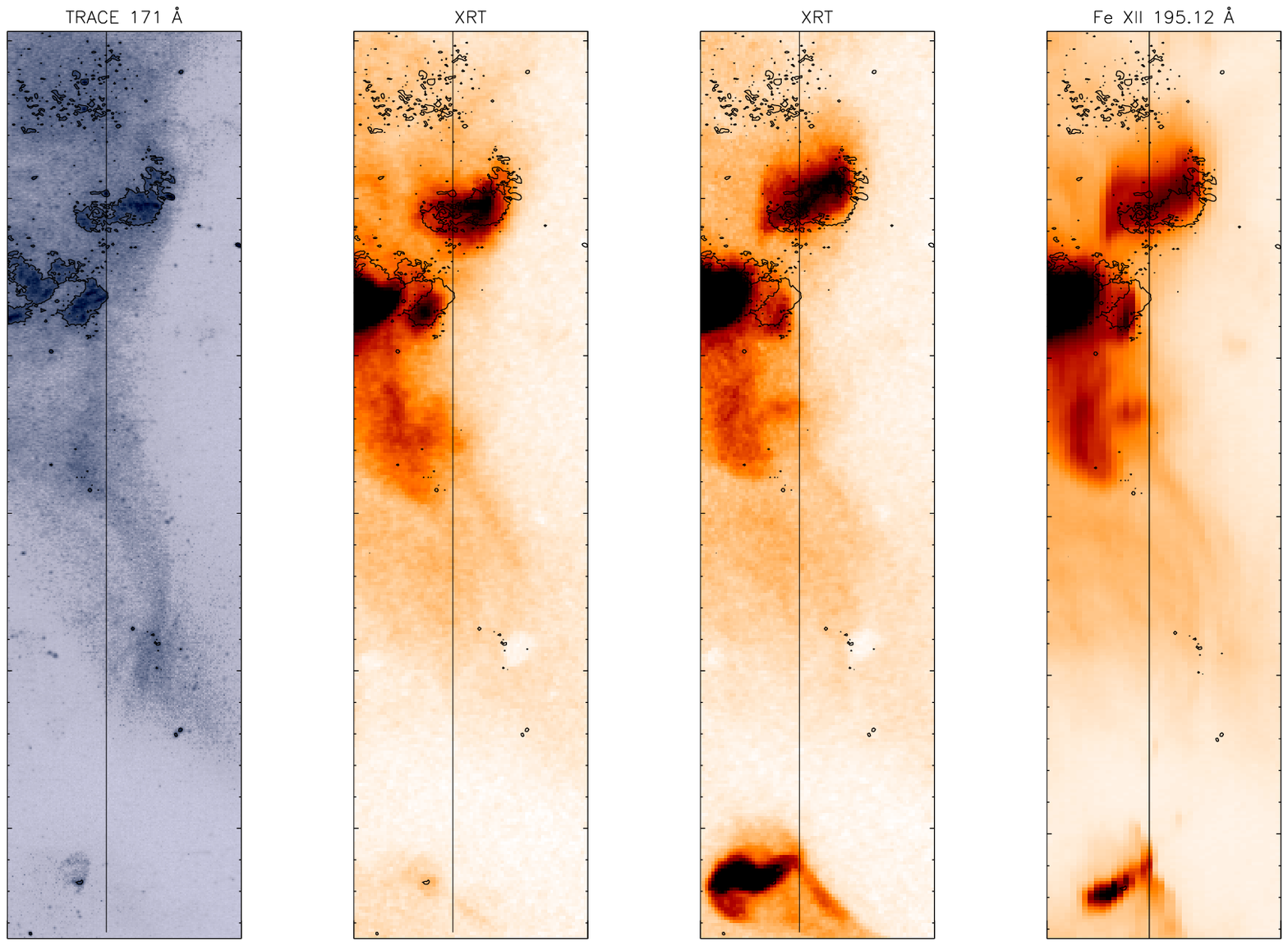}
\includegraphics[scale=0.7]{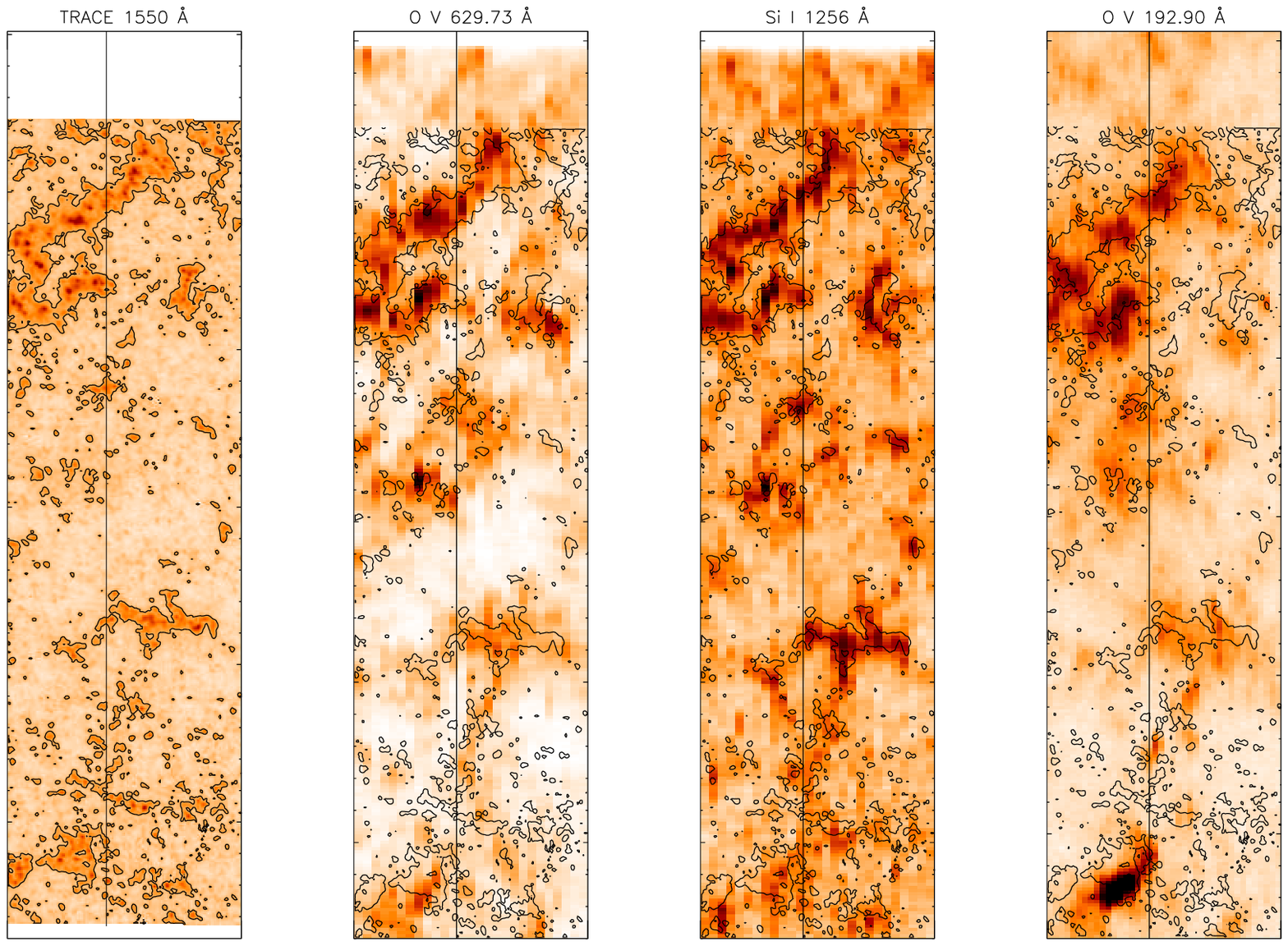}
\caption{{\bf Top, from left to right:}  TRACE 171~\AA\ image taken at 03:27~UT, XRT image at 03:27~UT, XRT image at 01:01~UT, EIS raster image in Fe~{\sc xii}~195.12~\AA\ obtained between 00:40~UT and 01:21~UT. All images have the TRACE 171~\AA\ contour and the SUMER slit position over-plotted. {\bf Bottom, from left to right:}
 TRACE 1550~\AA\ image taken at 03:28~UT, SUMER O~{\sc v} 629.73~\AA\ and S~{\sc i} 1256.49~\AA\ raster images taken between 02:43~UT and 02:56~UT, EIS O~{\sc v} 192.90~\AA\ raster taken from 00:40 to 01:21~UT. All images have the TRACE~1550~\AA\ contour and the SUMER slit position over-plotted. }\label{fig0}
\end{figure*}

We also calculated the electron density of the jet-like event and its footpoints which comprise the bright point and the micro-flaring site therein.  There are several limitations in estimating this quantity. We could obtain the electron density only for the region of lower dynamics which complies with the assumption of ionisation equilibrium. Therefore, the plasma density values are only reliable in the area enclosed with   a contour in Fig.~\ref{fig7}.   From the analysis of density sensitive line pairs available in our dataset, we concluded that the only line ratio which could provide reliable density diagnostics is the line ratio of Fe~{\sc xii} 186.88 to 195.12~\AA\ (the Fe~{\sc xii} 196.64~\AA\ was not observed).  The Fe~{\sc xiii} ratio of  203.82 to 202.04 is not suitable due to discrepancies in estimating densities above 10$^{10}$~cm$^{-3}$   \citep[for more details see][]{2009A&A...495..587Y}.  Equally, the blend  in the blue wing of the Fe~{\sc xiii} 203.82~\AA\ line makes the use of this line ratio quite uncertain  for events showing blue-shifts as in the present case.  

There are several blending issues concerning the Fe~{\sc xii} 186.88 and 195.12~\AA\ lines. A comprehensive study on this can be found in \citet{2009A&A...495..587Y}. The theoretical line ratio derived with  CHIANTI v6.0 includes for 186~\AA\ the two Fe~{\sc xii}  lines at 186.85~\AA\  and 186.88~\AA\ and for 195~\AA\ the two  Fe~{\sc xii} lines -- 195.12~\AA\ and 195.18~\AA. The Fe~{\sc xii} 186 is also blended by S~{\sc xi} 186.84 which was taken  as 19.5\% of S~{\sc xi}~191.27~\AA\ \citep{2009A&A...495..587Y}.  We obtained an average density value for the BP region of approximately  4 $\times$ 10$^9$~cm$^{-3}$, a  value of  around 4~$\times$~10$^8$~cm$^{-3}$ for  the surrounding corona  and a maximum value of 4~$\times$~10$^{10}$~cm$^{-3}$ for the density of the micro-flaring region at temperature 10$^{6.1}$~K which is the maximum of the formation temperature of Fe~{\sc xii}.


\section{Discussion and conclusions}

\subsection{On the physical parameters}
The physical parameters of X-ray jets were the subject of several studies \citep{2000ApJ...542.1100S, 2007PASJ...59S.751C, 2007PASJ...59S.771S,2010ApJ...710.1806D}. \citet{2000ApJ...542.1100S} derived the temperature and the density of X-ray jets from Yohkoh/SXT observations by analysing 16 jets and their footpoints. The jet temperatures were estimated in the order of 3--8~MK and the density in the range from 0.7 to 4.0 $\times$ 10$^9$~cm$^{-3}$. The authors found that the temperatures of the footpoint micro-flares were  similar to those of the jet, i.e. 4--8~MK, while the densities were slightly higher, e.g. 2.4--10.0 $\times$ 10$^9$~cm$^{-3}$. We should bare in mind, however, that physical parameters derived from imager observations can be uncertain. The first to deliver results on jet plasma quantities from Hinode observations were \citet{2007PASJ...59S.751C}. They used slot observations which can also give only a rough estimation of the physical quantities. For the two analysed jets, the temperatures ranged from 0.4~MK to 5.0~MK, while the density was 4 $\times$ 10$^9$~cm$^{-3}$.  The most recent published values by  \citet{2010ApJ...710.1806D} are 0.3--2.2~MK for the temperature and approximately 10$^9$~cm$^{-3}$ or less for the electron density. In comparison, we found in the footpoints of the jet, i.e. at the location of the micro-flare, a temperature (derived as the maximum of the formation temperature of the observed spectral line) of up to 12~MK while the density was 4~$\times$~10$^{10}$~cm$^{-3}$.  The average densities of the bright point where the outflow originates from, is 4~$\times$ 10$^9$~cm$^{-3}$ which is comparable to earlier studies. 

\subsection{Comparison with a 3D MHD model}

 We compared the observed dynamics and the physical parameters of the jet-like event analysed in the present paper with the most recent and  very close  to realistic  simulated jet-like event by  \citet{2008ApJ...673L.211M}. They  developed a 3D MHD model based on the interaction of  a twisted magnetic flux rope rising from the convection zone and expanding into the corona. A current sheet forms between the counter-aligned side of the flux tube with the ambient coronal field.  The authors describe the formation of a ``double-chambered'' vault shaped region  formed of current loops both emerged and reconnected.  The on-disk view of our jet does not permit us to make any comparison with these structures, but similar structures  were already described in the literature. In the simulated jet two regions of reconnected field lines are formed: first are the reconnected open field lines which is comparable to what we  see shortly after the first energy deposition  took place, followed by closed loops rising from below. In comparison with the simulated feature, we clearly distinguish the formation of a high velocity outflow along open magnetic field lines followed very shortly by the expulsion of numerous loops which, however, is not seen in the simulations. Recently it has been found  by \citet{2010arXiv1002.1675S} that  X-ray jets are always associated with pre-existing or newly emerging (at X-ray temperatures) coronal bright points. They concluded that a necessary condition for the generation of outflows is the interaction of closed and open magnetic field lines. Therefore, there are always  pre-existing loop structures, i.e. coronal bright points, that are expelled during the event.  There is also the possibility that some of the expelled loops  represent newly reconnected loops but we strongly believe that most of the observed loop structures   lifting away belong to the pre-existing bright point. One strong argument for this is the fact that after the jet-like event the BP almost fully disappears at X-ray temperatures. A careful check on the lower temperature images (TRACE and EUVI Fe~{\sc ix/x} 171~\AA) including cooler spectral lines (transition region temperatures) indicated that there was still a remaining lower lying BP which consequently  produced another jet but only at low temperatures until the BP fully disappeared.  This scenario was observed in the majority of the BP producing jets in \citet{2010arXiv1002.1675S} but it will be a subject of a forthcoming article.

Can  other  mechanism(s)  rather than flux emergence  trigger magnetic reconnection in BPs?  The fact that jet-like events occur predominantly  at coronal hole boundaries  and inside coronal holes where closed loop structures exist, and  that coronal holes are rotating quasi-rigidly in respect to the surrounding quiet Sun \citep{1975SoPh...42..135T}, suggest that  there exist the necessary conditions these two topologically different magnetic fluxes to be pushed together to form a current sheet  and reconnect.  We also detected for the first time several energy depositions during the phenomenon which indicates several episodes of reconnection may have taken place.  This scenario needs to be explored theoretically.   

 The physical parameters obtained from  the Moreno-Instertis \etal experiment  compare very well with the results of the present study.
The temperature of the reconnection site of T $\approx$ 3 $\times$ 10$^7$~K is very close to the 1.3 $\times$ 10$^7$~K derived here. The high plasma temperature was, however, not detected in the outflowing plasma, but that could be simply due to a low signal-to-noise in the observed line as discussed earlier. Both the velocity of the plasma and the duration of the event are quite alike.

As we mentioned earlier the event described here was selected from a large number of events registered during a co-ordinated campaign of Hinode and SoHO. A forthcoming paper will give details on the dynamics and physical parameters of large number of events identified by \citet{2010arXiv1002.1675S}.

\begin{acknowledgements} The author thanks ISSI, Bern for the support of the team ``Small-scale transient phenomena and their contribution to coronal heating''. The author  thanks K. Galsgaard and J.G. Doyle for the careful reading and useful comments of this manuscript. Research at Armagh Observatory is grant-aided by the N.~Ireland Department of Culture, Arts and Leasure. We also thank STFC for support via grants ST/F001843/1 and PP/E002242/1. Hinode is a Japanese mission developed and launched by ISAS/JAXA, with NAOJ as domestic partner and NASA and STFC (UK) as international partners. It is operated by these agencies in co-operation with ESA and NSC (Norway).

\end{acknowledgements}

\bibliographystyle{aa}


\begin{figure*}[!h]
\center
\includegraphics{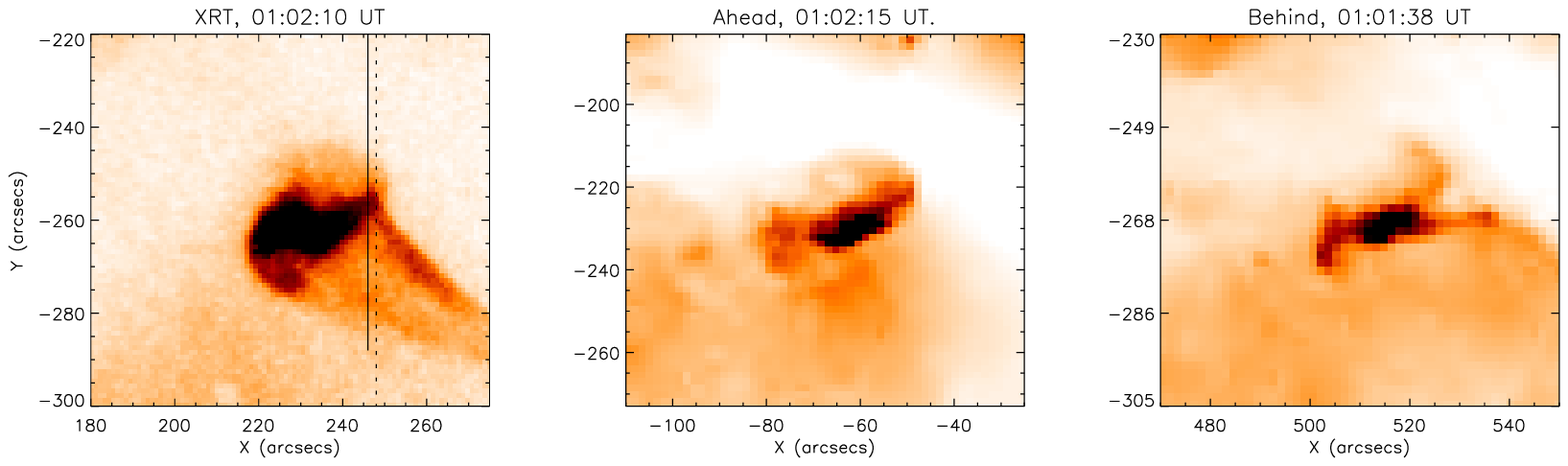}
\includegraphics[scale=1.]{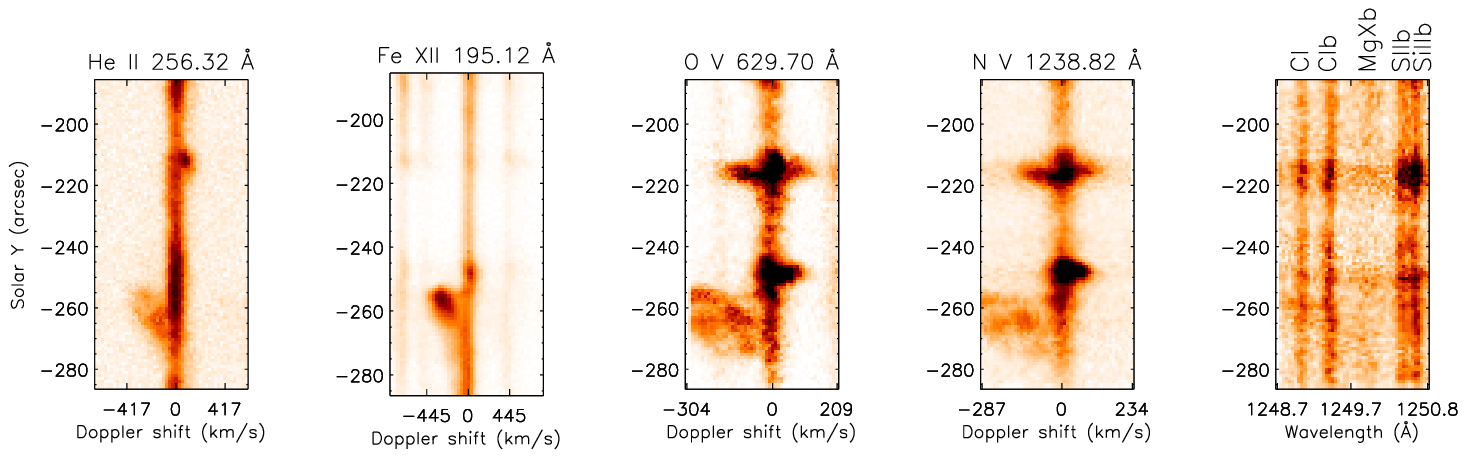}
\caption{{\bf Top:} XRT image with overplotted  the SUMER slit position, EUVI A and  B images during the jet-like phenomenon. {\bf Bottom (from left to right):}
 EIS He~{\sc ii} 256.32~\AA\ and Fe~{\sc xii} 195.12~\AA\  spectra along the EIS slit at 01:01:42~UT; SUMER O~{\sc v} and N~{\sc v} spectra along the SUMER slit; and a 
 spectral window comprising C~{\sc i} 1248.0~\AA, C~{\sc i} 1249.0~\AA, Mg~{\sc x} 624.9~\AA\ taken in second order, Si~{\sc ii} 1250.41~\AA\ and 
 S~{\sc ii} 1250.58~\AA (for details on blends and formation temperatures consult Table~\ref{T1} ) at 01:01:19~UT.}
\label{fig1a}
\end{figure*}


\begin{figure*}[!ht]
\center
\includegraphics{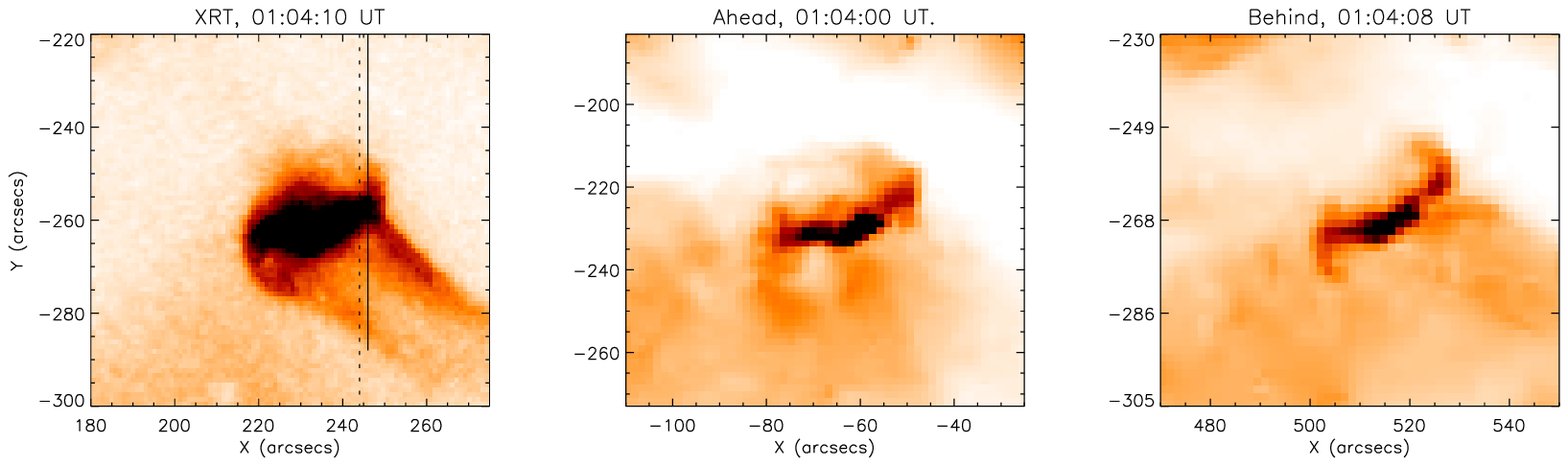}
\includegraphics[scale=1]{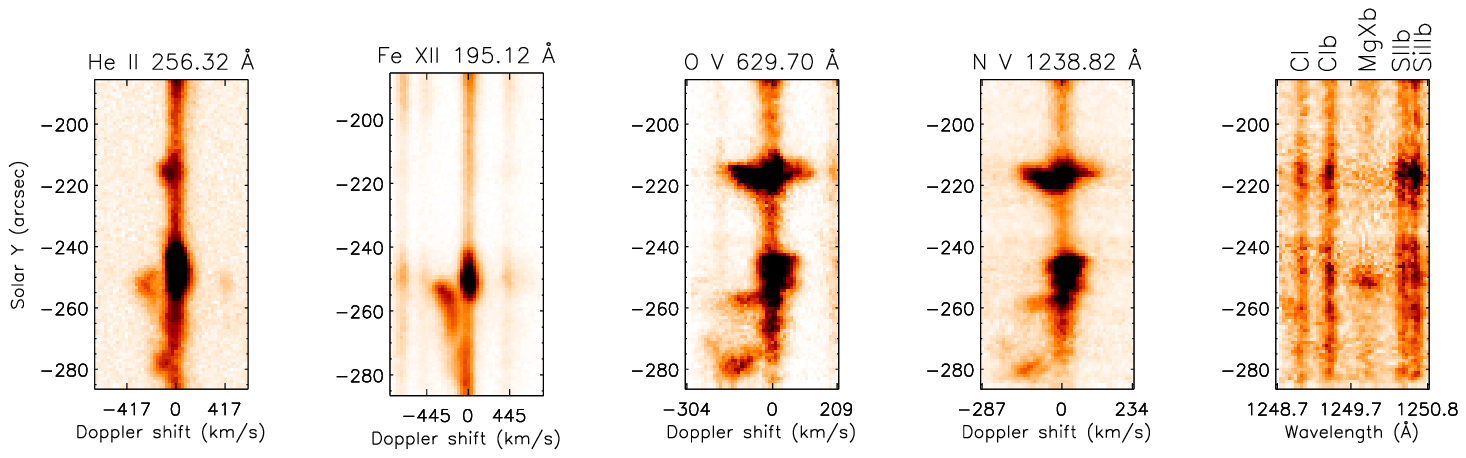}
\caption{The same as Figure~\ref{fig1a} with EIS at 01:03:46~UT and SUMER at 01:03:26~UT.}
\label{fig2a}
\end{figure*}

\begin{figure*}[!ht]
\center
\includegraphics{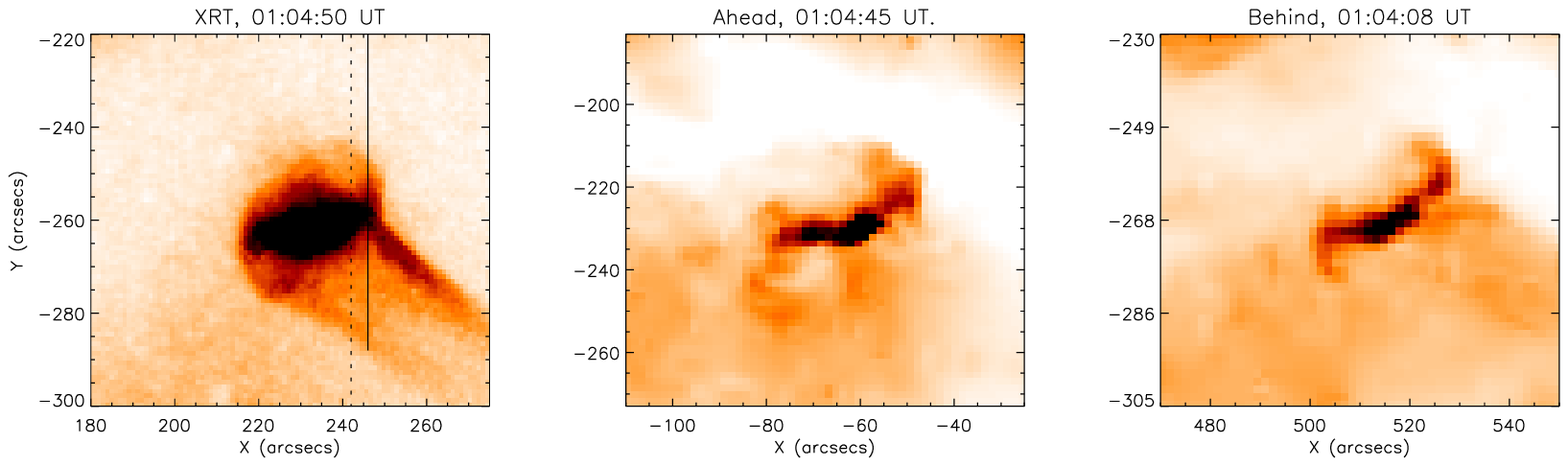}
\includegraphics[scale=1.]{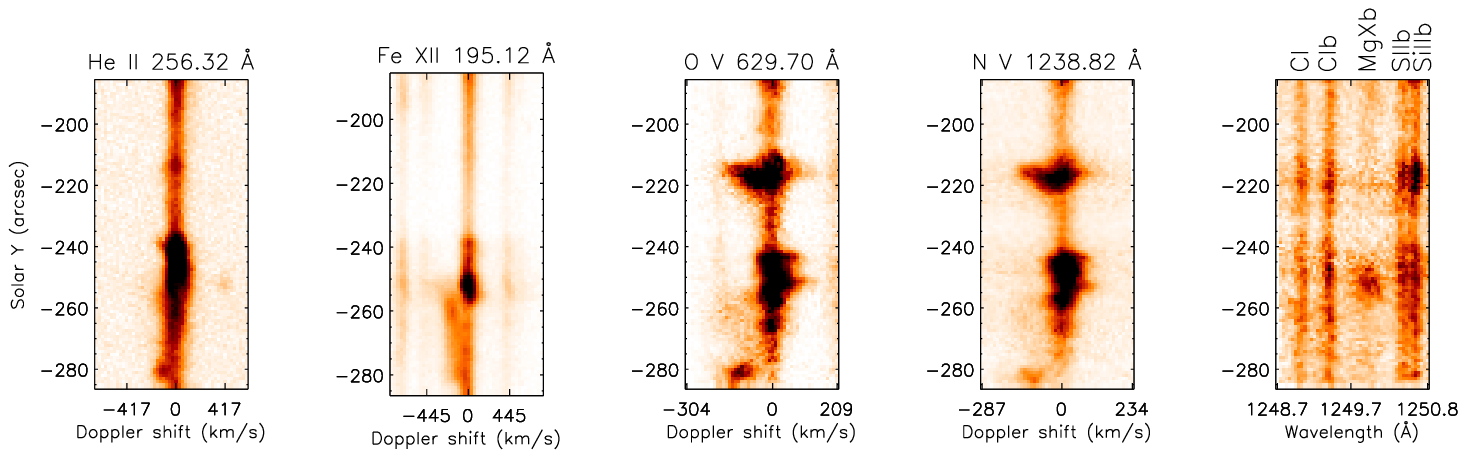}
\caption{The same as Figure~\ref{fig1a} with EIS  at 01:04:48~UT and SUMER at 01:04:26~UT.}
\label{fig3a}
\end{figure*}

\begin{figure*}[!ht]
\center
\includegraphics{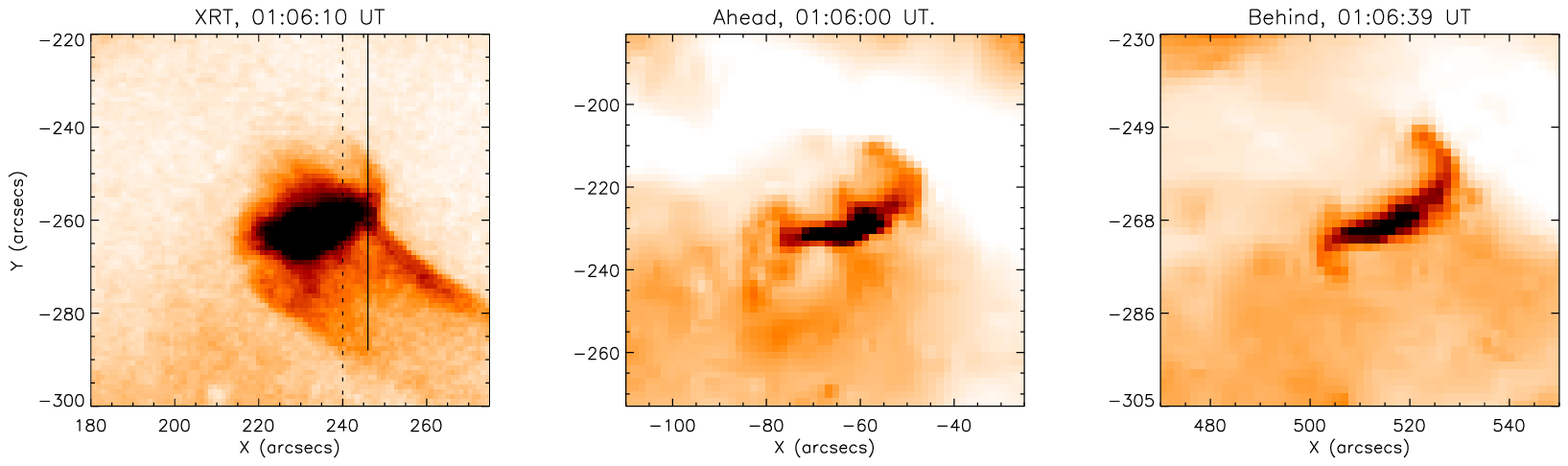}
\includegraphics[scale=1.]{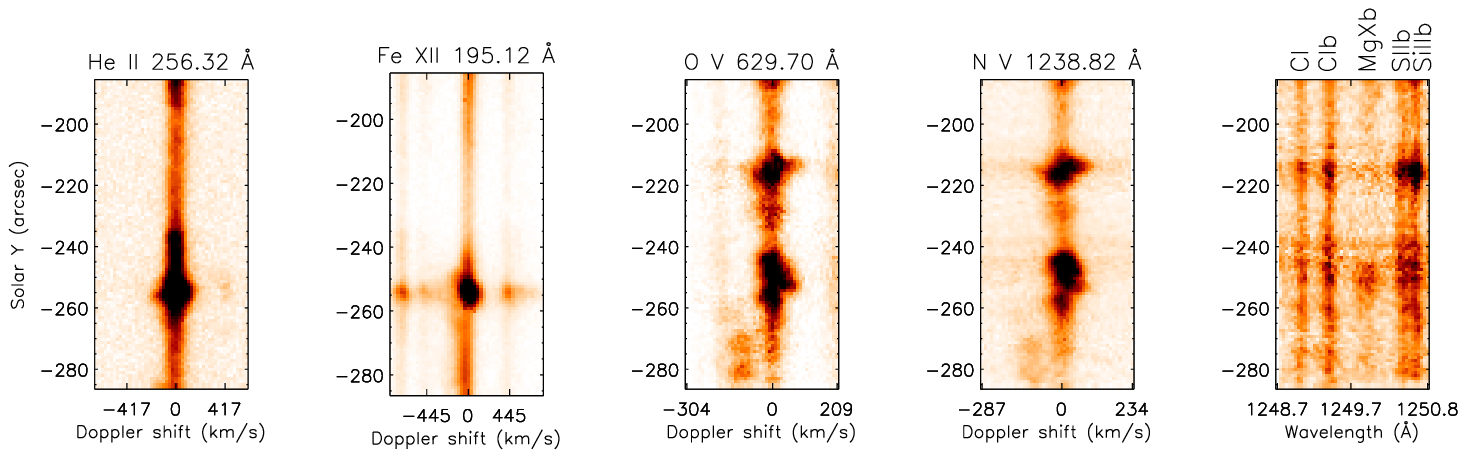}
\caption{The same as Figure~\ref{fig1a} with EIS at 01:05:50~UT and SUMER at 01:05:32~UT.}
\label{fig4a}
\end{figure*}

\begin{figure*}[!ht]
\center
\includegraphics{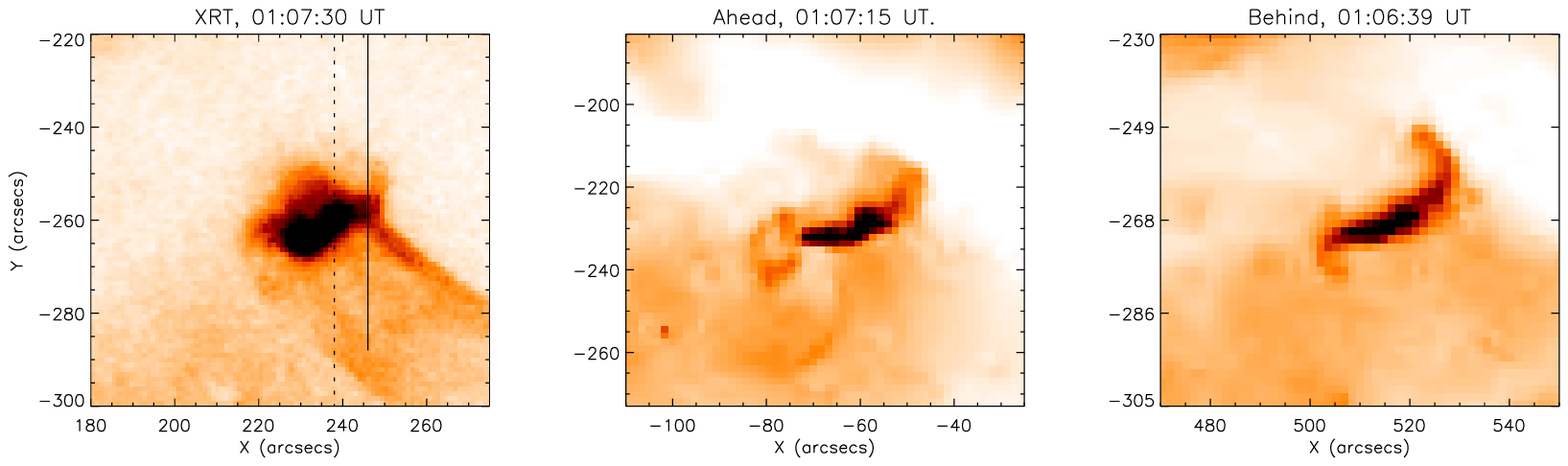}
\includegraphics[scale=1.]{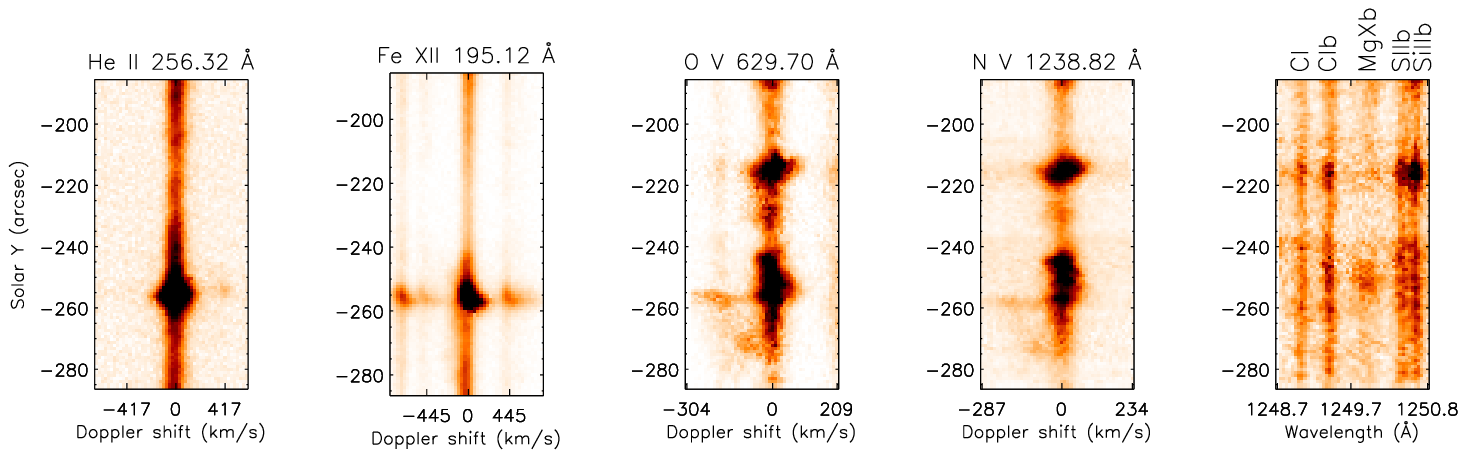}
\caption{The same as Figure~\ref{fig1a} with EIS at 01:06:52~UT and SUMER at 01:06:32~UT.}
\label{fig5a}
\end{figure*}


\begin{figure*}[!ht]
\center
\includegraphics{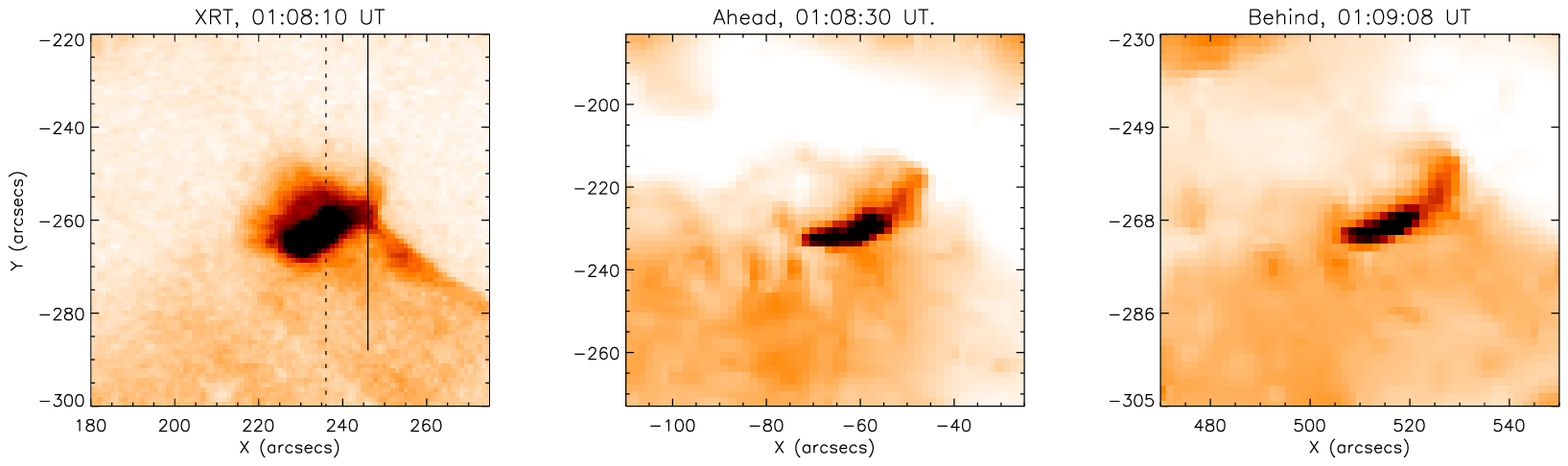}
\includegraphics[scale=1.]{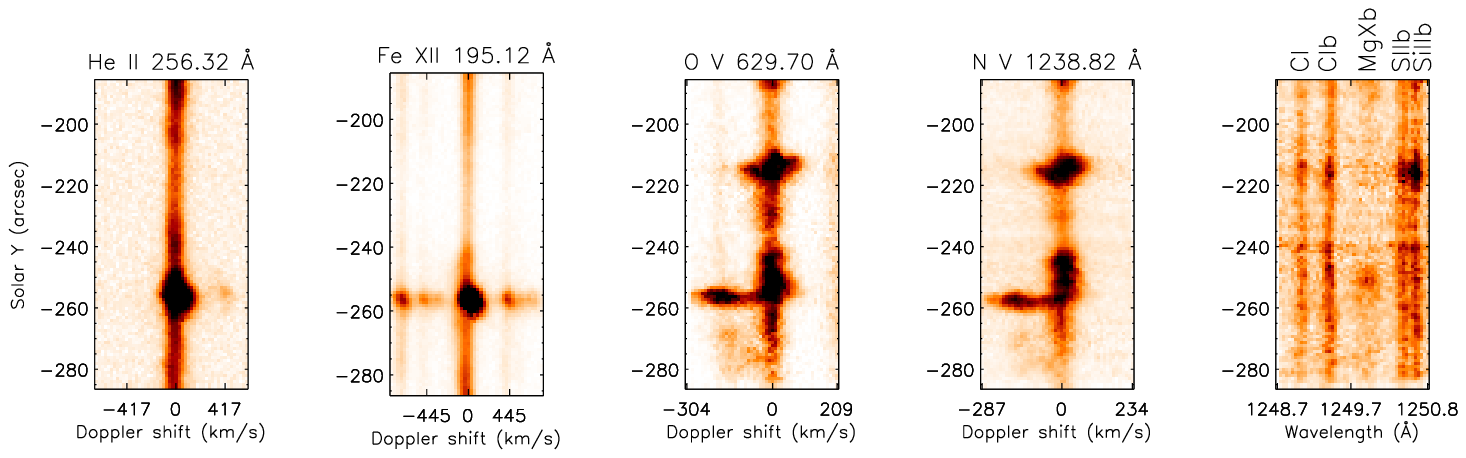}
\caption{The same as Figure~\ref{fig1a}. with EIS at 01:07:54~UT and SUMER at 01:07:40~UT.}
\label{fig6a}
\end{figure*}

\end{document}